\documentclass[a4paper, 12pt]{article}
\bgroup

\usepackage[utf8]{inputenc}%utf8 no funciona en windows quizas y hay que usar T1 o latin1
\usepackage[english]{babel}    %agrega terminos en castellano y seperaciòn en palabras
\usepackage{hyperref}        %Las notas al pie y los titulos en la intro se hacen hipervinculos
\usepackage{graphicx}         %Paquete de graficos y figuras 
\usepackage{float}          %Paquete que mueve figuras
\usepackage[usenames]{color}  %Nombres de los colores
\usepackage{anysize}		%Cualquier margen
\marginsize{2cm}{2cm}{2cm}{2cm}  %Margen cheto	
\usepackage{verbatim}		%Esto no me acuerdo qué era
\usepackage[justification=centering]{caption} 	%Centrar todas las figuras
\usepackage{soul} %Es para resaltar con el comando \hl
\usepackage{blkarray} % Para generar las matrices con filas y columnas comentadas
\usepackage{float} % Para fijar las figuras
\usepackage[rightcaption]{sidecap}
\usepackage{setspace}
\usepackage{authblk}
\onehalfspace
\usepackage{multirow}
\usepackage{hyperref}
\hypersetup{
    colorlinks=false,
    linkcolor=blue,
    filecolor=magenta,      
    urlcolor=blue,
}
\usepackage{blkarray}
\usepackage{lineno}
\usepackage{multirow}
\usepackage{diagbox}
\usepackage{slashbox}
\title{News-sharing on Twitter reveals emergent fragmentation of media agenda and persistent polarization}
\date{\today}

\author[1,3]{Tomas Cicchini}
\author[1,2]{Sofia M. del Pozo}
\author[1,2]{Enzo Tagliazucchi}
\author[1,2]{Pablo Balenzuela \thanks{balen@df.uba.ar}}

\affil[1]{\small{Departamento de F\'isica, Facultad de Ciencias Exactas y Naturales, Universidad de Buenos Aires. Av.Cantilo s/n, Pabell\'on I, Ciudad Universitaria, C1428EGA, Buenos Aires, Argentina.}}
\affil[2]{\small{Instituto de F\'isica de Buenos Aires (IFIBA), CONICET. Av.Cantilo s/n, Pabell\'on I, Ciudad Universitaria, C1428EGA, Buenos Aires, Argentina.}}
\affil[3]{\small{Instituto del C\'alculo (IC), UBA-CONICET. Intendente Güiraldes 2160, Ciudad Universitaria, Pabellón II, 2do. piso, C1428EGA, Buenos Aires, Argentina.}}

\begin{document}
\maketitle

\begin{abstract}
News sharing on social networks reveals how information disseminates among users. This process, constrained by user preferences and social ties, plays a key role in the  formation of public opinion. In this work we study news sharing of main Argentinian media outlets in Twitter, using bipartite news-user networks, in order to understand if the emergence of affinity groups is driven by the underlying political polarization. We compare the results between an electoral and non-electoral year and between a set of politically active users and a control group. We found that users' behavior produces well differentiated  communities of news articles identified by a unique distribution of media outlets in all analyzed datasets. In particular, these communities split into two groups which reflect the dominant ideological polarization in Argentina. We also found that users form  two well differentiated groups identified by their preferences in media outlets consumption. These two groups of media outlets display a bias towards the two main political parties that rule the political life in Argentina.  These results reveal consistently that ideological polarization is the main driving force shaping the Argentinian news sharing in Twitter.
\end{abstract}
\section*{Introduction}

\par The mass media play a preponderant role in the formation of public opinion \cite{McCombs1972,Guo2015}. Nowadays, many people are exposed to news and share them  through social media \cite{weaver, noticias_en_Twitter_1, noticias_en_Twitter_2}. Therefore, understanding the way in which information circulates through them plays a fundamental role in the process of opinion formation \cite{Feezell2018,Gilardi2021}.

\par The circulation of information is strongly influenced by how users connect among them in different social media. This connectivity \cite{DelVicario554} arises with formation of ties based on affinity, group membership or trust in influential individuals or organizations, among other causes \cite{cita_homofilia_1, cita_homofilia_2}. As these ties emerge, social networks become clustered, leading to constraints on information flow.

\par The emergence of highly connected groups of individuals is a topological feature that repeatedly arises in studies of social networks in relation to discussions around specific topics. They have been observed in networks defined by preferential message propagation (\emph{retweet networks}, in the case of Twitter) \cite{Menczer2011,Becatti2019} , as well as in networks of followers \cite{Cinelli2021}. These groups reflect the clustering of individuals based on different measures of similarity among users \cite{Cota2019, Cinelli2021, Vicario_2016, Baronchelli_2021, Makse_2021, Zollo_2017} creating homogeneous communities that are frequently known as \emph{echo chambers} \cite{jamieson2008echo,quattrociocchi2016echo,gilbert2009blogs}.

\par In these works, the reported polarization phenomena refers to groups which extreme their opinions on discussions around a specific topic (gun control, vaccination, etc.). However, topics are rarely discussed  in isolation \cite{Baumann_2021} and the phenomenon of issue alignment phenomena plays \cite{Baldassarri_2008} a key role in polarization in the political process, leading to antagonistic ideological states. 

\par The role of social media and news sharing in the formation of ideological echo chambers has become an important topic of research in recent years. For instance, in \cite{Bail2018} the authors analyze if exposure to partisan information among Twitter users contributes to the formation of ideological echo chambers. In \cite{Bakshy2015} they analyze how social media, such as Facebook, shapes the exposure to ideologically diverse news. 
In \cite{Aruguete2020} they retrieve measures of cognitive congruence (ideology) in the phenomena of news sharing in data from Brazil, Argentina and USA.

\par In this paper we investigate which are the key features leading to the emergence of well defined groups in the process of news sharing of the main media outlets in Argentina. The main hypothesis of this work is that the way in which users share news on a social networks, such as Twitter, is mediated by personal preferences and ideological affinity in such a way that it is possible to detect emerging groups as a consequence of these interactions. Under this hypothesis, we sought to provide a quantitative answer to these questions:

\begin{itemize}
    \item Are news sharing constrained by features related to users or news? Or does the information diffuse freely on social media?
    \item Why are certain groups of news more frequently co-shared between them than with others?
    \item Do users tend to form clusters according to their given preferences in news consumption?
    \item Can the news consumption profile of users be used to define echo-chambers?
    \item Does media consumption in social media reflect the political polarization in Argentina? Does this depend on whether users explicitly share their political preferences, or on whether data is collected during an electoral campaign?
\end{itemize}

To answer these questions, we first present a brief \nameref{Bck} of the media outlet and political landscape of Argentina. Then, we summarize the \nameref{DyM} with which the analysis of this work was performed. The results of these analysis can be found in section \nameref{Res}, where users and news projections are treated separately. Finally, the discussions that emerge from these analyses are presented in section \nameref{Dis}.

\section*{Background}\label{Bck}

\par In this section we sketched the political and media landscape in Argentina during the analyzed period, in order to contextualize the results obtained in this manuscript.

\par National elections in Argentina have two mandatory instances: the primary election, called PASO (in Spanish: \textit{Primarias, Abiertas, Simultáneas y Obligatorias}; in English: Open, Simultaneous and Obligatory Primaries), and a general election. In 2019 these instances took place on August 11th and October 27th, respectively. If needed, depending on the results of the general elections, a third instance, called \textit{ballotage}, can also take place.

\par In the last ten years, the political landscape in Argentina was dominated by two coalitions: a center-left coalition led by Cristina Fernandez de Kirchner (called {\it Frente de Todos}) and a center-right one led by Mauricio Macri (called {\it Juntos por el Cambio}). Cristina Kirchner was the Argentinian president in the periods $2007-2011$ and $2011-2015$ and Mauricio Macri  in  $2015-2019$ \cite{cantamutto2016}.
\par In the 2019 elections, the candidates of the center-left coalition were  {\it  Alberto Fernandez} and  {\it Cristina Fernandez de Kirchner}, while  {\it Mauricio Macri} competed for his re-election as president for the center-right coalition with  {\it Miguel Angel Pichetto} as candidate for vice-president. Also  {\it Axel Kicilliof} and  {\it Maria Eugenia Vidal} were candidates to Governor of the Buenos Aires province for the {\it Frente de Todos} and the {\it Juntos por el Cambio} coalitions, respectively.

\par Concerning the media landscape, the ranking of the most consumed digital media outlets in Argentina is dominated by three players: {\it Infobae}, {\it Clarin} and {\it La Nacion} with approximately 20 millions of unique users each as of in 2020 \cite{todomedios} according to Comscore data. They are followed by a second group of  outlets with audiences oscillating between 6 millions and 13 millions of unique visitors. Among them, we can highlight {\it Pagina 12}, {\it Ambito Financiero}, {\it TN Noticias} and {\it El Destape Web}.

\par Ideological bias of some of them has been reported, as for instance in \cite{bonner2018}, where {\it Pagina 12} is classified as a left-of-center broadsheet newspaper, {\it Clarin} as centrist tabloid with the highest circulation of the national newspaper and {\it La Nacion} as a elite, center-right, broadsheet. Also in \cite{mitchlstein2017}, authors refer to {\it La Nacion} as a newspaper which has historically adopted an elitist stance marked by favoring the interests of the upper classes and to {\it Clarin} as a newspaper with a centrist ideology, together with a general interest perspective and a broad, middle-class target audience.

\par During 2008-2014 there was a confrontation between the government of {\it Cristina Fernandez de Kirchner} and mass media  \cite{mitchlstein2017}. It resulted in the emergence of a group of newspapers close to the \textit{Kirchner} government’s policy ({\it Pagina 12} among them), and a group of newspapers whose editorials were strongly critical of all measures taken by the administration  at this time ({\it Clarin}, {\it La Nacion} and others) \cite{mitchlstein2017, Becerra2012, Yeager2014}.

\section*{Data and Methods}\label{DyM}

Twitter users sharing links to media content were selected in order to analyze the emergence of structures from their complex pattern interactions, following a similar approach than in  \cite{weaver}.

\subsection*{Data}

\par Twitter data was acquired using the social network official API \cite{api_url}, together with custom developed python codes. The acquisition process consisted of the following steps:

\begin{enumerate}
    \item \textbf{User selection: live download.} Twitter activity was downloaded using the Stream Listener tool in the Twitter API. A set of politically active users were filtered by keywords (discarding the retweets) associated with politicians, electoral alliances and political parties (see keywords in Supplementary Information) during the primary 2019 Argentinian elections (between August $5^{th}$ and August $12^{th}$). Also, a control group was selected by keywords associated to the name of the main media outlets in Argentina (discarding also the retweets) during the period from August $29^{th}$ to September $30^{th}$ 2019 and from June $4^{th}$ to July $4^{th}$ 2020 (see keywords in SI). We collected the same amount of users for both datasets: $38K$ for 2019 and $35K$ for 2020.
    
     \item \textbf{Twitter activity: tweets download} Full twitter activity from both datasets were then downloaded during two different periods of time: from August $29^{th}$ to September $30^{th}$ 2019 and from June $4^{th}$ to July $4^{th}$ 2020. We collected 1368914 tweets of politically active users in 2019 and 987271 tweets in 2020. The control dataset was 511308 tweets in 2019 and 576137 tweets in 2020. 

    \item \textbf{Embedded \emph{URLs} tweets}: From previously downloaded tweets,  we keep only those containing embedded \emph{URLs}. 
 %The identified tweets were downloaded using the Trend Line tool from the Twitter API,
 
    \item \textbf{URL filter}: Outlet domains were obtained from the \emph{URLs} together with an Argentinian media data base provided by ABYZ News Links \cite{abyz}. We only keep  tweets from the major twenty Argentinian media outlets, obtaining 80811 and 66688 tweets for the politically active users dataset and 31811 and 41593 tweets for the control dataset in 2019 and 2020 respectively.
\end{enumerate}

The final bipartite network data is shown in Table \ref{bipartite_table}.

\begin{table}[h!]
\centering
\begin{tabular}{|l|r|r|r|r|}

\hline
\textbf{Dataset}               & \multicolumn{1}{l|}{\textbf{Period}} & \multicolumn{1}{l|}{\textbf{\# Tweets}} & \multicolumn{1}{l|}{\textbf{\# Urls}} & \multicolumn{1}{l|}{\textbf{\# Users}} \\ \hline
\multirow{2}{*}{Politically active users}          & 2019                               & 80810                                   & 40188                                 & 10748                                  \\ \cline{2-5} 
                               & 2020                               & 66687                                   & 34580                                 & 10115                                  \\ \hline
\multirow{2}{*}{Control group} & 2019                               & 31810                                   & 14604                                 & 5158                                   \\ \cline{2-5} 
                               & 2020                               & 41437                                   & 17285                                 & 6756                                   \\ \hline

\end{tabular}
\caption{\textit{Bipartite networks description of politically active users and control group datasets.}}
\label{bipartite_table}
\end{table}

\subsection*{Methods}

\subsubsection*{Bipartite Networks and their projections}

\par The complex pattern of news shared by multiple users can be mapped onto bipartite networks  following the procedure sketched in \cite{weaver}.

\par  Bipartite networks have two different classes of nodes and  can be projected into news and user layers. Connections in the news projection indicate co-consumption across users, while the user projection describes users connected by news in common.

\par Projections of bipartite networks can be done in several ways, as for instance following the method developed in \cite{Saracco_2017}. Understanding that projections induce several noisy edges between nodes of the same layer, such a method proposes a null model to define which edge plays a significant role on the monopartite network structure. Based on this idea, here we follow a similar approach, combining a simple projection method \cite{NewmanProjection} with a significance filter introduced in \cite{significancia}. As in \cite{weaver}, a hyperbolic projection was used in order to mitigate the influence of highly connected nodes of both layers (see SI 2). The significance filter keep those links whose weights are meaningful in comparison with those expected from a null stochastic model where nodes keep their degree and the total strength. Same approach was followed in \cite{Silva_diputados}.

\par Given that these projections do not necessarily produce fully connected networks, we kept the largest connected components for further analysis. 

\subsubsection*{Networks metrics}

We mainly focused on the analysis of collective structures and the role of nodes within these structures.

\par To detect communities in both projections of the bipartite networks we used a Python implementation of the Louvain algorithm \cite{louvain, python_louvain} based on the optimization of the modularity $Q$. Due to the stochastic nature of the Louvain algorithm, the obtained community partitions may differ from each other in a comparatively small number of nodes. To obtain a well-defined membership metric of the nodes to the given communities, we constructed consensus networks \cite{redconsenso}, allowing the robust assignment of nodes to communities. 

\par Even though limitations of the community detection based on modularity for weighted networks are known \cite{matando_modularity}, the results obtained here were robust when compared with other algorithms like Infomap and Label Propagation, as can be seen in SI 4, where normalized mutual information was computed. 

\par We also analyzed  the role of users in the network by means of the participation coefficient and the within module degree as were defined in \cite{cartography}. Given a partition in communities of a network, the within module degree basically measures how connected is a node within their own community and the participation coefficient how well is connected a node with other communities.

\subsubsection*{Nodes metrics}

To analyze the properties of emergent structures on both projections of the bipartite networks, we focus on metrics related to the semantic content and the media outlet membership (for the news projection) and to the user profile of media consumption (for the user projection).

\textbf{News projection}

When nodes correspond to media content, two features are relevant for the analysis performed here: the media outlet in which they were published and their semantic content.

\begin{itemize}

 \item {\bf Media outlet} We first classify each news article according to the media outlet where it has been published.

 \item {\bf Semantic Content Analysis} The following steps were applied to the text of each news article:

\begin{enumerate}
\item \textbf{tokenization}: each element of the corpus was separated in individual terms, and non-alphanumeric characters and punctuation was removed. All terms were converted to lowercase.
\item \textbf{stopwords filtering}: using the Spanish stopwords database provided by nltk \cite{nltk}, the most common (and thus the least informative) words were filtered out.
\item \textbf{term basis construction}: a term basis was generated from the set of used terms. 
\item \textbf{frequency description}: each news article was described as a term frequency vector with entries given by the basis computed in the previous step, i.e., the i-th term corresponded to the number of times the i-th term of the basis appeared in each news.
\item \textbf{tf-idf description}: to mitigate bias due to excessive contribution of frequently used words and increase the contribution of unusual (but informative) words, the term frequency - inverse document frequency (tf-idf) statistic was computed \cite{NGUYEN201495}, resulting in the following value for the i-th element of the vector news representation:

\begin{equation}
        v_{i} = f_{i} \cdot \log \left(\frac{N}{N_{i}}\right)
\end{equation}
where $N$ is the number of documents in the corpus and $N_i$ is the number of documents where the i-th term appears.
\end{enumerate}

\par After these processing steps, the news corpus was described as a matrix $M \in \mathrm{R}^{n \times m}$, with $n$ the number of documents in the corpus and $m$ the number of basis terms.

 \item {\bf Unsupervised topic detection} Starting from this mathematical representation of the corpus it is possible to detect the main topics (i.e. groups of similar articles with roughly the same semantic content) by performing Non-negative Matrix Factorization (NMF) \cite{Lee1999}\cite{documentClusteringNMF} on the document-term matrix (M). NMF results in the factorization of the news-terms matrix $M$ as the product between two matrices:
\begin{equation}
    M \approx N \cdot W \textit{ , con } N \in \mathrm{R}^{n \times t}\textit{ y }W \in \mathrm{R}^{t \times m}
\end{equation}
where $t$ is the chosen number of topics, and $N$ and $W$ are the news in topics dimension and the topics in terms dimension, respectively. Both matrices are composed only of positive entries, allowing their simple interpretation.

\par Following the procedure sketched in \cite{PaperSeba}, we define the media agenda as the fraction of articles belonging to each topic.

 \item {\bf Sentimental Content Analysis} Sentiment analysis were performed in phrases of each news article were candidates of the main political coalitions were mentioned. We use the algorithm developed in \cite{perez2021pysentimiento}. This algorithm assigns a label $+1,0,-1$ according if the sentiment of the sentences is positive, neutral or negative respectively. With this information, each news article will have phrases with negative, neutral and positive phrases towards the mentioned candidates. Then we apply a metric called Sentiment Bias Statistic as defined in \cite{Albanese_2020} using the sentiment algorithm classifier. The procedure is the following:

\begin{enumerate}
\item In each text, we detected phrases mentioning the names of the candidates of the two main coalition disputing the national elections in Argentina 2019 (``Cristina", ``Kirchner", ``Alberto", ``Fernandez" from the center-left one and  ``Mauricio", ``Macri" or ``Picheto" from the center right).
\item We applied the sentiment analysis for these sentences and counted the amount of positive, negative, and neutral mentions for each of the candidates. 
\item For each news article,  $\#KF_+$ ($\#KF_-$) stands for fraction of positive (negative) mentions of \textit{Cristina Kirchner} or \textit{Alberto Fernandez} and $\#MP_+$ ($\#MP_+$) for positive (negative) mentions of \textit{Mauricio Macri} or \textit{Miguel Angel Picheto} and the Sentiment Bias statistic $SB$ is defined according to equation \ref{eq:SB}.  
\end{enumerate}

\begin{equation}
    SB = (\#KF_+ - \#KF_-) - (\#MP_+ - \#MP_-)
    \label{eq:SB}
\end{equation}

The  Sentiment Bias statistic $SB$ is a measure of the  bias towards one of the coalitions. If $SB>0$, the bias is positive towards the candidates of the center-left one (\textit{Cristina Kirchner} or \textit{Alberto Fernandez}) compared with the candidates of the center-right coalition (\textit{Mauricio Macri} or \textit{Miguel Angel Picheto}). If $SB<0$ is the same towards the center-right coalition.

\end{itemize}

\textbf{User projection}\label{users_properties}

\par Users are linked by the news they consume. We quantified the diversity in the media outlets consumed by each user in terms of the following vector: 

\begin{equation}
    m^i = (1,...,j,...,0)
\label{eq:md}
\end{equation}

where the $m^i_j$ component indicates the number of news from the $j$ media outlet shared by the $i$ user. Given the heterogeneity in the distribution of news outlets in the corpus, we  introduce a corrected version of this metric \footnote{The subindex $c$ stands for \emph{corrected}}:

\begin{equation}
\label{eqn:corrected_medium_vector}
(m^i_{c})_j = m^i_j \cdot \log \left(\frac{N}{N_j}\right),
\end{equation}

where $N$ is the total number of users and $N_j$ the total number of users sharing news  belonging to the $j$ media outlet. Here, the factor $\log \left(\frac{N}{N_j}\right)$ corrects the potential bias caused by a given article being shared by multiple users.

Given that users can share news from multiple outlets or, conversely, share news from only one of them, we estimated how diverse was the user behavior in terms of the  shared media outlets using the maximum value of the consumed media vector. This measures the Lack of Diversity (LD) in the user behavior:

\begin{equation}\label{pol}
    LD_i = \max_{j \in M}\{(m^i_{c})_j\}
\end{equation}

where M is the total number of media outlets in our data set. After a normalization, this lack of diversity lies between 0 and 1.

\section*{Results}\label{Res}

\par Our results aim to provide insight about the underlying mechanism in the formation of tight communities of news articles and users. Our hypotheses are that individual preferences and ideological leaning should give rise to the emergence of communities identified either to the media outlets or their semantic content. 
%These results should be consistent in both projections and across all available datasets. 

\par Here we show the results corresponding to the datasets of politically active users in 2019 and 2020. Table \ref{table_data} shows the final number of nodes for each period. The same table related to the control group can be found in SI 3, in addition with a full topological summary.

\begin{table}[h!]
\begin{center}
\begin{tabular}{ | c| c| c | c | } 
\hline
\textbf{Period} & \textbf{Users} & \textbf{News articles} \\ \hline
29/07 to 29/08 2019 & 3625 & 12221 \\ \hline 
4/06 to 4/06 2020 & 4323 & 10975 \\ \hline
\end{tabular}
\caption{\textit{Number of nodes in the users and news networks of the politically active users dataset.}}
\label{table_data}
\end{center}
\end{table}

\par
\begin{figure}[h!]
    \centering
    \includegraphics[scale = .5]{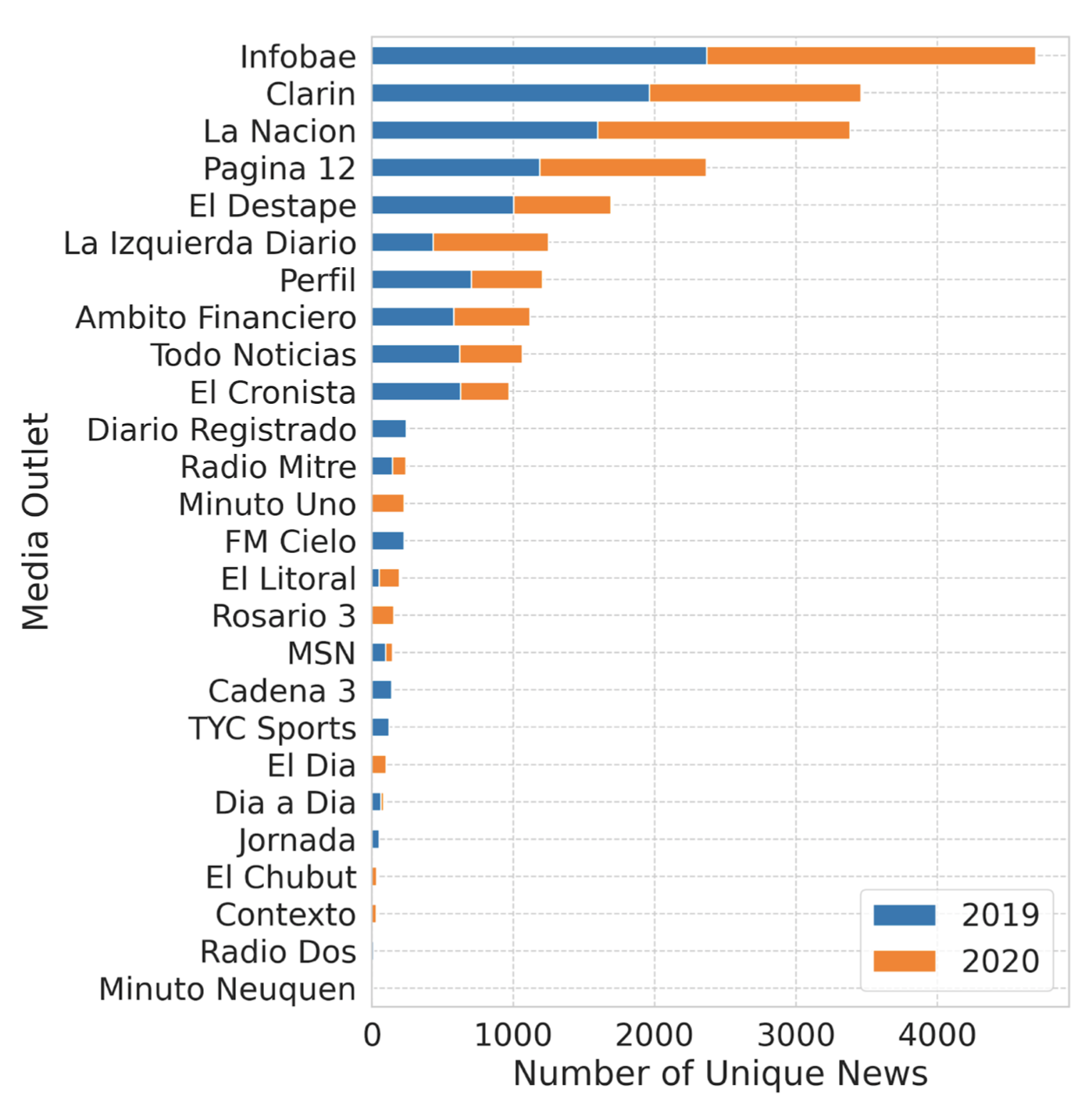}
    \caption{\textit{Distribution of news articles according to media outlet}. Here we plot the amount of articles of each media outlet corresponding to both analyzed years in the set of politically active users.}
    \label{Fig0_DistriMedios}
\end{figure}

 In Fig. \ref{Fig0_DistriMedios} we show the distribution of news articles classified by media outlets for 2019 and 2020. We can see that distribution of scrapped news is similar to the level of consumption in Argentina of the main media outlets \cite{todomedios}.

\par In what follows, we analyze networks emerging from both projections of the bipartite users-news networks.

\subsection*{The news projection}

\par In this projection, nodes represent news articles and edges are proportional to the number of users co-sharing the corresponding news. We performed a topological analysis of these networks, together with the semantic analysis of news content.

\par
\begin{figure}[h!]
    \centering
    \includegraphics[scale = .4]{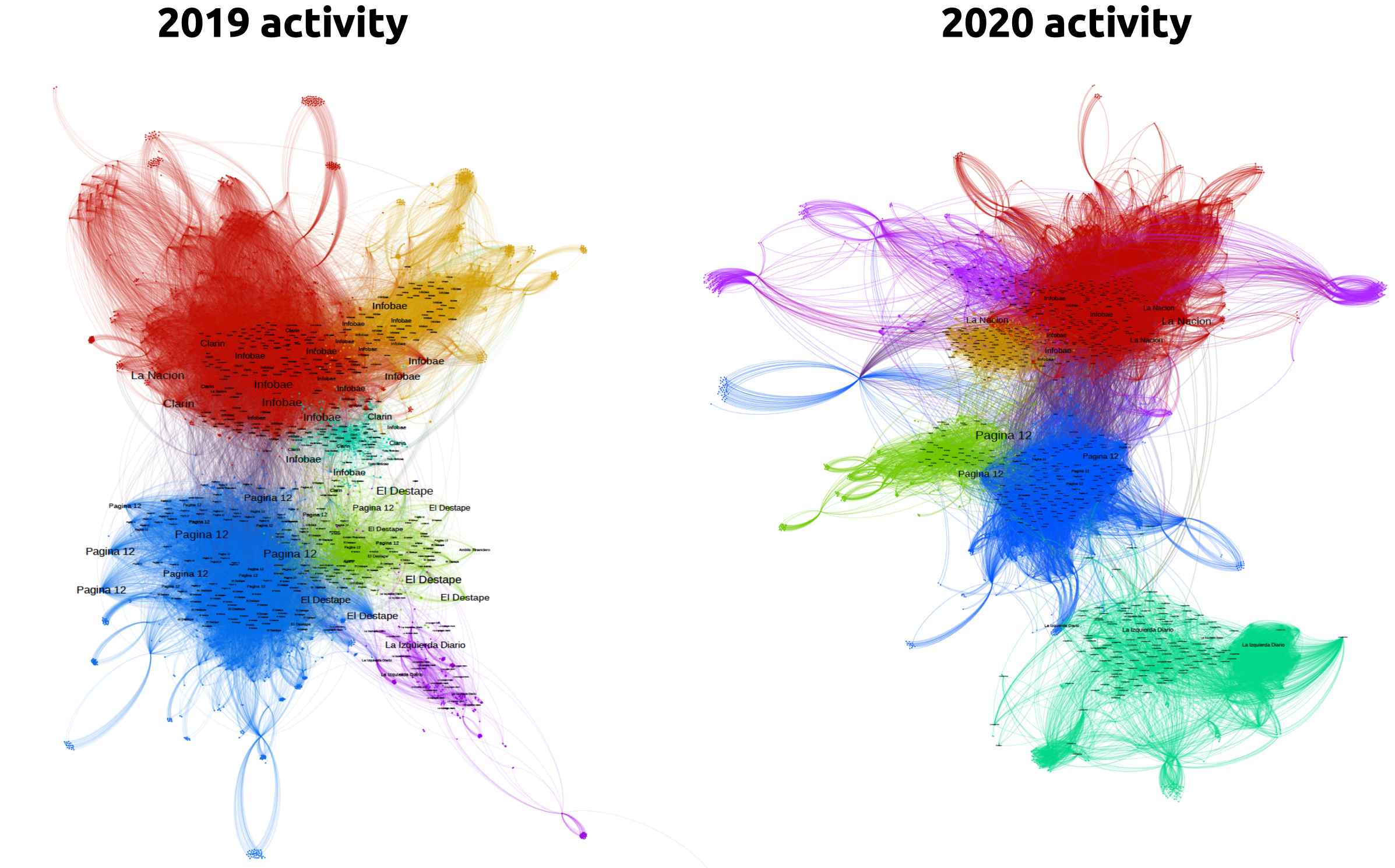}
    \caption{\textit{2019 and 2020 news networks visualization. Nodes were colored by community membership and labeled by media outlet proportionally to their within-module degree}.}
    \label{Fig1_RedesNoticias} 
\end{figure}

 We performed community detection in both data sets, yielding the structures visualized in Fig. \ref{Fig1_RedesNoticias}. The displayed labels correspond to the media outlets of the news with greater within-module degree. Colors represent the different communities. It is possible to appreciate the strong relationship between the news media outlets, the emerging communities structure and how it persists across both datasets: \textit{Pagina 12} and \textit{El Destape Web} from one side, \textit{La Nacion},  \textit{Clarin} and \textit{Infobae} from the other. 

\par
\begin{figure}[h!]
    \centering
    \includegraphics[scale = .35]{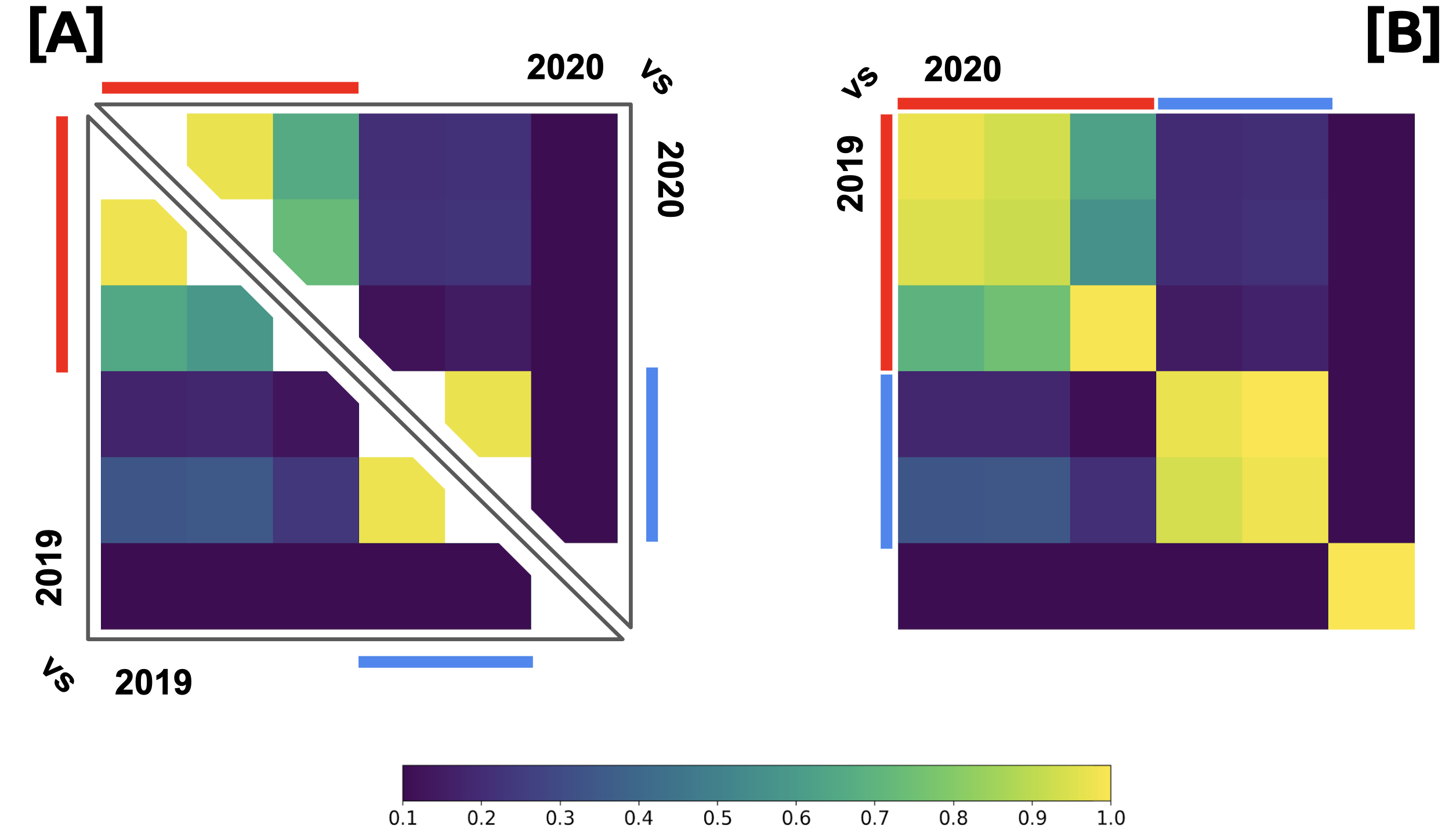}
    \caption{\textit{Consistency of news network. Panel [A] accounts for similarities among communities of same years and Panel [B] compares 2019 against 2020.}}
    \label{Fig2_comparacion_medios}
\end{figure}

  In order to quantify the role of media outlets in the emergence of this structure, we describe each community by an array $C^{m}_i$ where each component accounts for the number of news of the media $i$ in community $c$.  Then we calculate the cosine similarities among these vectors. The results are shown in  Fig.\ref{Fig2_comparacion_medios}. Panel [A] accounts for similarities among communities of the same years and Panel [B] compares 2019 against 2020. Three groups arise and are consistent in the comparison between both years: a first one composed by three communities, the second one by two communities and a  small one in third place. 

\par The media outlet vector distributions shows a strong polarization between the two main  groups of news: the first one is composed with news from  mainly three outlets: {\it Clarín}, {\it La Nación} and {\it Infobae} and the second one is composed basically with news from {\it Página 12} and {\it El Destape}. The third group in size is a small community with news from {\it La Izquierda Diario} ( the newspaper of the socialist worker party). 

\par
\begin{figure}[h!]
    \centering
    \includegraphics[scale = .55]{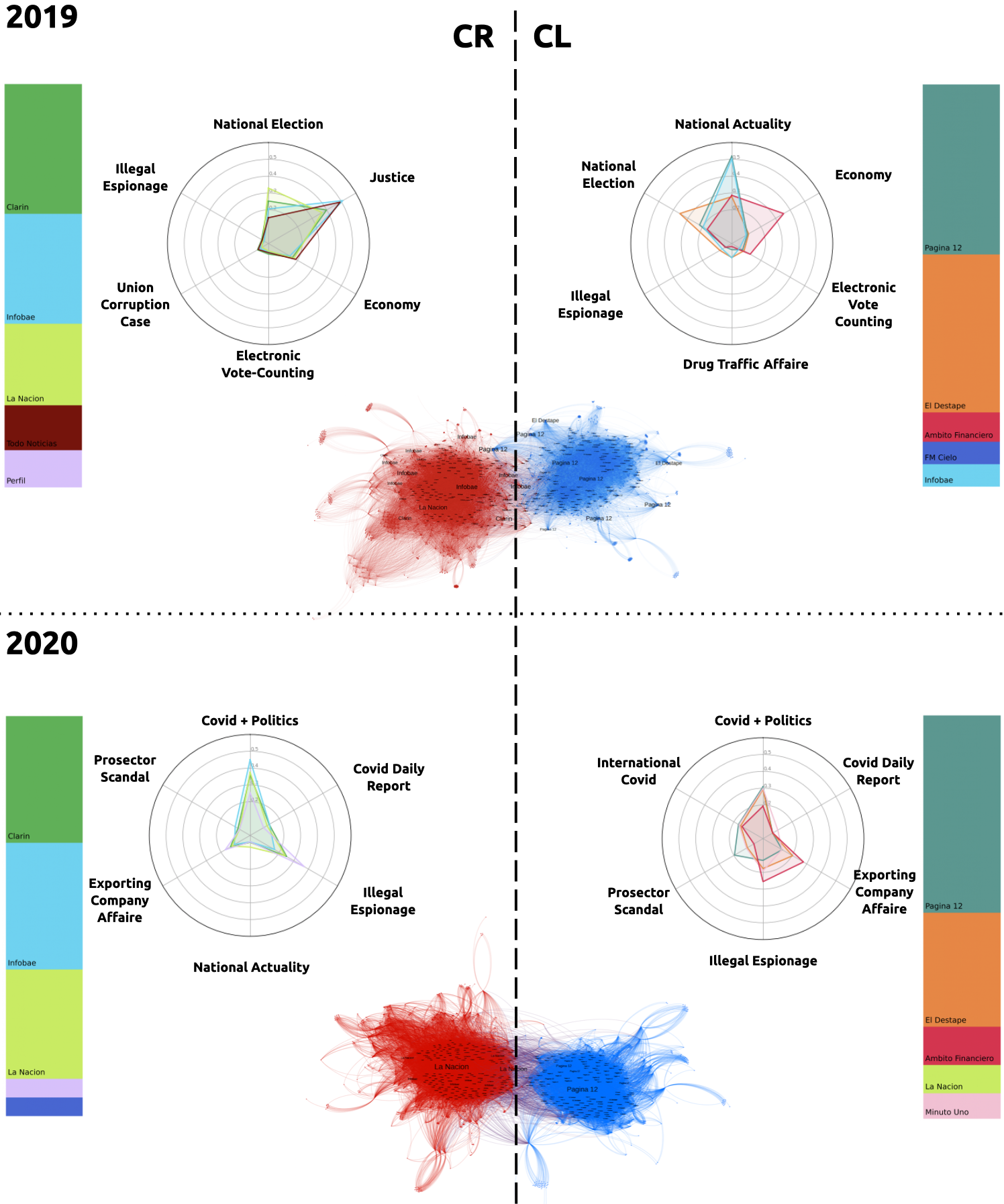}
    \caption{\textit{Media outlets distributions and topic decomposition for the 2019 and 2020 two main communities. The stacked bars represent the media outlet distribution, while the radar plots  display the media agenda. The agenda of each Figureoutlet is indicated with lines colored with the same color as in the stacked bar.}}
    \label{Fig3_Comunas1920}
\end{figure}

 We complement this analysis by performing topic decomposition of the news content within each community (see Methods). Information is detailed in Table \ref{NewsTable} for the six main communities of each data set and plot a more detailed version in Fig. \ref{Fig3_Comunas1920} for the two biggest communities.

\begin{table}[h!]
\begin{center}
\small
\begin{tabular}{ | p{0.125\linewidth}| p{0.09\linewidth} | p{0.25\linewidth} | p{0.35\linewidth} |} 
\hline
\multicolumn{4}{|c|}{\textbf{2019}} \\ \hline
\textbf{Community Alias} & \textbf{Nodes (\%)} & \textbf{Main Media oulets} & \textbf{Main Topics} \\ \hline
 Center-Right I & 18.68 & Clarin, Infobae, La Nacion & Justice, National Elections, Economy
\\ \hline
Center-Left I & 14.75 & Pagina 12, El Destape & National Actuality, Economy, National Elections \\ \hline
Center-Left II & 8.34  & El Destape, Pagina 12 &  National Actuality, National Election\\ \hline
International & 5.5 & Infobae & International Actuality \\ \hline
Center-Right II & 4.4 & Clarin, Infobae, La Nacion, Todo Noticias& Justice, National Election \\ \hline
Radical Left & 3.4 & Izquierda Diario & Politics, Public Health, Economy \\ \hline
%\textbf{Year} & \textbf{Community Label} & \textbf{Community Alias} & \textbf{Nodes (\%)} & \textbf{Main Mediums} & \textbf{Main Topics} \\ \hline
\multicolumn{4}{|c|}{\textbf{2020}} \\ \hline
\textbf{Community Alias} & \textbf{Nodes (\%)} & \textbf{Main Media oulets} & \textbf{Main Topics} \\ \hline
Center-Right I  & 16.07 & Clarin, Infobae, La Nacion & Covid+Politics, Illegal Espionage \\ \hline
Center-Left I & 11.33& El Destape, Pagina 12 & Covid + Politics, Exporting Company Affaire, Illegal Espionage \\ \hline
Center-Left II & 8.61 & Pagina 12, El Destape & National Actuality, Economy, Covid Daily Report \\ \hline
Center-Right II & 7.43 & Infobae, La Nacion, Clarin & International Covid, Covid Daily Report, Justice \\ \hline
Radical Left & 7.1 &La Izquierda Diario &  llegal Espionage, Globally Known Racial Issue \\ \hline
International & 4.5 & Infobae &  International Actuality \\ \hline

\end{tabular}
\caption{\textit{Main communities description for 2019 and 2020 news networks}}
\label{NewsTable}
\end{center}
\end{table}

\par In Fig.\ref{Fig3_Comunas1920}, we focus the analysis on the two main communities of both years. Here, media outlet distributions are shown in the stacked bars and the media agendas in the topic space in the radar plots. These results set up the idea that the meanwhile media outlet distribution remains in time, media content is informative of what is happening in the time span where news were analyzed. 

\par According to the media landscape described in section Background, the observed distribution of media outlets in two main groups  parallels the two most voted and widespread political coalitions in 2019 national Argentinian elections. While {\it Página 12} and {\it El Destape} are known to support the center-left political party, {\it Clarín} and {\it La Nación} are known for being the  ``opposition media'' to it \cite{mitchlstein2017, Becerra2012, Yeager2014}.

\par In light of these results we wonder if it is possible to define a metric which accounts for the support of these different groups of media outlets to the mentioned candidates in the analyzed dataset. First, we noticed that in the topic  \textbf{National Elections}, present in the two main communities of 2019 dataset, the names of the main candidates are highlighted  (see word clouds for each community in Supplementary Information): $Cristina$, $Alberto$ and $Fernandez$  from the Center-Left coalition and $Vidal$ and $Macri$ from the Center-Right one. Second, we can use sentiment analysis to phrases mentioning the candidates in media content to measure positive or negative bias towards them, as described in section Methods.

\par We calculate Sentiment Bias (SB) for all media content in two main groups of communities for 2019 and 2020. The center-left one, dominated by {\it Página 12} and {\it El Destape}, gives $SB = 0.17 \pm 0.05 $ in 2019 and  $SB =0.02 \pm 0.05$  in 2020. For the center-right one, dominated by {\it Clarín}, {\it La Nación} and {\it Infobae}, we have $SB =-0.04 \pm 0.05$ in 2019 and  $SB =  -0.13 \pm 0.05$ in 2020. Sentiment bias values are shown in table  \ref{SBbig_table} 

\begin{table}[h!]
\centering
\begin{tabular}{|lll|}
\hline
\multicolumn{3}{|c|}{\textbf{Sentiment Bias}}                                                                                    \\ \hline
\multicolumn{1}{|l|}{\backslashbox{Communities}{Year}} & \multicolumn{1}{r|}{2019}                      & \multicolumn{1}{r|}{2020} \\ \hline
\multicolumn{1}{|l|}{Center-left group}                            & \multicolumn{1}{l|}{0.17 $\pm$ 0.05  \textbf{*}} & 0.02 $\pm$ 0.05             \\ \hline
\multicolumn{1}{|l|}{Center-right group}                            & \multicolumn{1}{l|}{-0.04 $\pm$ 0.05}            & -0.13 $\pm$ 0.05 \textbf{*} \\ \hline
\end{tabular}
\caption{\textit{Sentiment bias values of the sum of all Center-left and Center-right communities of news networks in both years. Bold asterisks denote that sentiment bias values are significantly different from zero with $p_{value} <0.01$.} }
\label{SBbig_table}
\end{table}

\par The first observation is that the difference between the Sentiment Bias of the two main groups persists in time. We quantified the statistical significance of these differences between them in two ways. Firstly we computed the Wilcoxon rank-sum test \cite{python_ranksums}. For 2019 and for 2020 the hypothesis was rejected at $1\%$ significance level ($p_{value}<0.01$). Second, we generated mean value distributions by bootstrapping with repositioning the sentiment bias sets. The null hypothesis was that both distributions had the same mean value. Like as the Wilcoxon test, we obtained $p_{value}<0.01$ for both years. 

\par The second test if the obtained values are significantly different from zero. 
Here, we can observe a clear difference between both years: On one hand, in 2019 $SB>0$ ($p<0.01$) in the community dominated by {\it Página 12} and {\it El Destape} (Bias towards \textit{Kirchner-Fernandez} or against \textit{Macri})  and no significant bias in the community dominated by {\it Clarín}, {\it La Nación} and {\it Infobae}. Let's keep in mind that 2019 was the year when Alberto Fernandez and Cristina Kirchner won the elections. On the other hand, in 2020, $SB>0$ ($p<0.01$) in the community dominated by {\it Clarín}, {\it La Nación} and {\it Infobae} (Bias towards Macri or against Kirchner-Fernandez) and no significative bias in the community dominated by {\it Página 12} and {\it El Destape}.  So, these results suggest a  displacement in the support to the  center-left coalition in 2019 to the center-right in 2020, when Fernandez and Kirchner where the incumbent president and vice-president during the SARS-COV-2 pandemic year 2020.

\subsection*{The user projection}

\par In this projection, nodes represent users and edges weights are proportional to the number of news co-shared by these users. In the previous section we found that the identity of the emergent structures was mainly driven by the distribution of media outlets shared in each community. Here, we complement the topological analysis of networks in this projection with a description of each user by its media consumer profile given by the corrected media distribution defined in Eq. \ref{eqn:corrected_medium_vector}.

\par
\begin{figure}[h!]
 \centering
    \includegraphics[scale = .5]{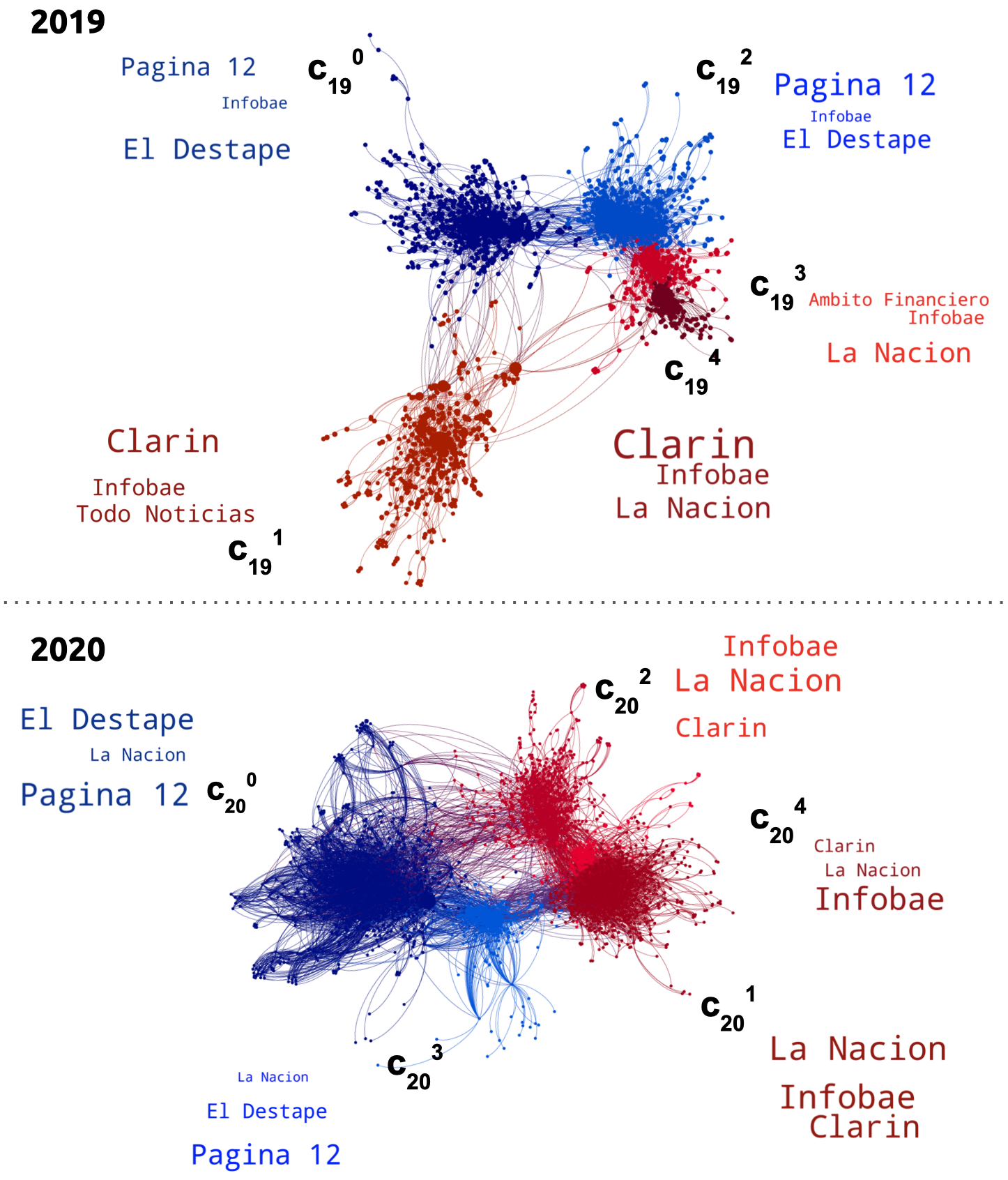}
    \caption{\textit{2019 and 2020 user networks visualization. Nodes were colored by communities membership. Word clouds display the averaged corrected media distribution of each community.}}
    \label{Fig4_Red_Usuarios19y20}
\end{figure}

 We performed community detection in both years' datasets. In Fig. \ref{Fig4_Red_Usuarios19y20} we can see highly structured users networks for the 2019 and 2020 user networks. Nodes were colored according to their community membership (keeping red tones for center-right leaning outlets and blue tones for center-left ones) and the labels correspond to  the average communities media vectors, according to Eq. \ref{eqn:corrected_medium_vector} ($<m^{i}_{c}>$). In Table \ref{tabla_usuarios} we describe the relative sizes and the main media outlet vectors for the five biggest communities. We can observe also here two different groups of communities: one with users which, in average, read news from {\it Página 12} and {\it El Destape} and others which read news from {\it Clarín}, {\it La Nación} and {\it Infobae}.

\begin{table}[h!]
    \centering
    \begin{tabular}{ | p{0.13\linewidth} | p{0.4\linewidth} |} 
\hline
\textbf{Nodes (\%)} & \textbf{Main Consumed Media Outlets} \\ \hline
\multicolumn{2}{|c|}{\textbf{2019}} \\ 
\hline
         15.86 & El Destape, Pagina 12\\ \hline
         12.61 & Clarin, Todo Noticias, Infobae, La Nacion\\ \hline
         11.09 & Pagina 12, El Destape \\ \hline
         5.58 & La Nacion, Infobae, Ambito Financiero, Pagina 12\\ \hline
         5.31 & Clarin, La Nacion, Infobae\\ \hline
    \multicolumn{2}{|c|}{\textbf{2020}} \\ \hline

         16 &Pagina 12, El Destape\\ \hline
         11.17 & La Nacion, Infobae, Clarin \\ \hline
         8.11 & La Nacion, Infobae, Clarin \\ \hline
         7.35 & Pagina 12, El Destape \\ \hline
         7.12 & Infobae \\ \hline
    \end{tabular}
    \caption{\textit{Main communities features for 2019 and 2020 users networks.}}
    \label{tabla_usuarios}
\end{table}

\par This visualization suggests that users form also well defined groups as those observed in previous works and they are strongly connected by the media they share in Twitter. So the question we ask these datasets is the following: Do they form homogeneous and well differentiated communities in terms of media consumer profile?

\par
\begin{figure}[h!]
 \centering
    \includegraphics[scale = .35]{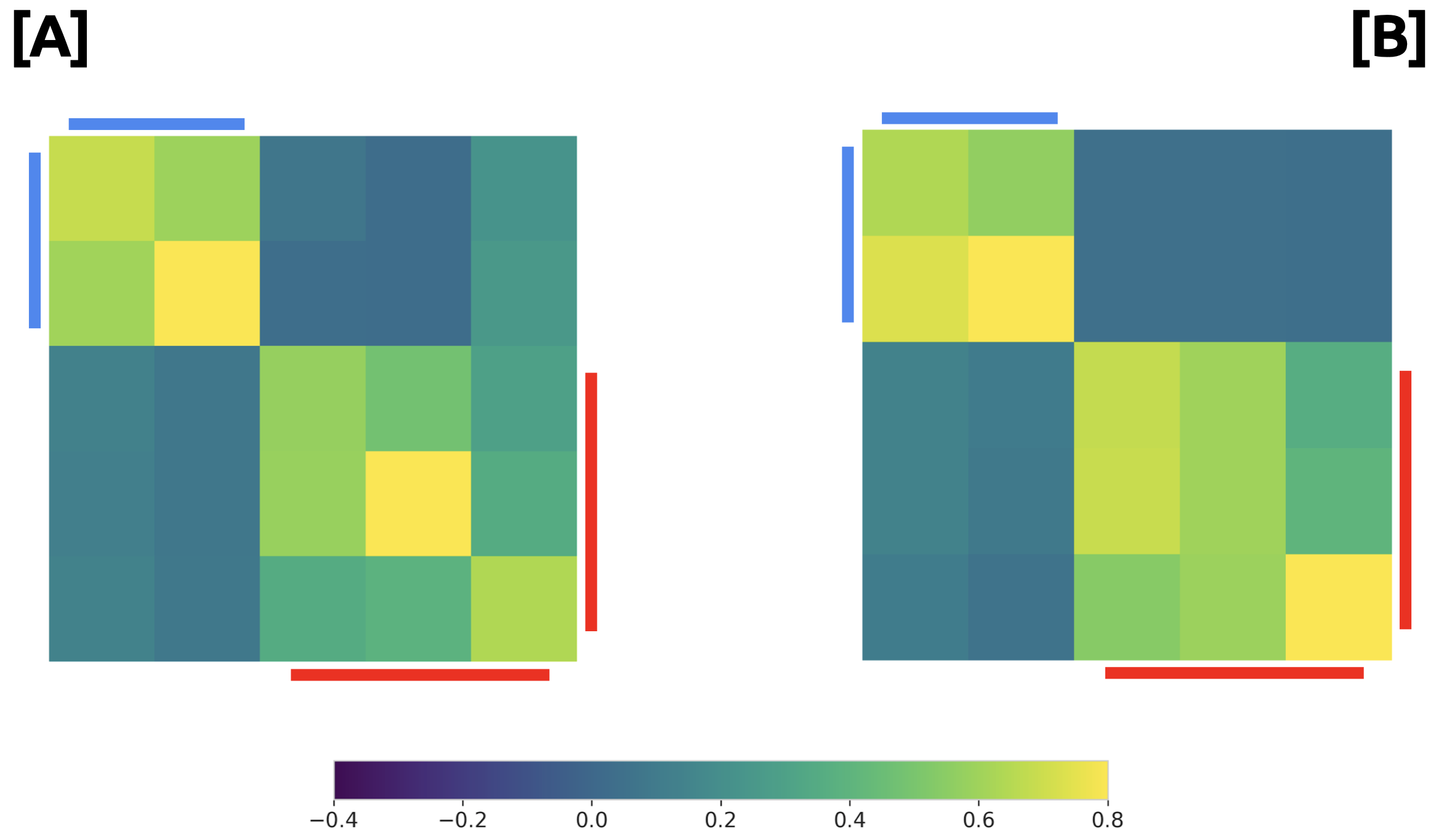}
    \caption{\textit{Similarities between users and average communities in media-consumed vectors. [A] and [B] accounts for 2019 and 2020 data sets, respectively. The i,j-th element of each figure corresponds to compute de median of the cosine similarities distribution between the average media-consumed vector of the i-th community and all the users media-consumed vectors belonging to the j-th community.}}
    \label{Fig5_Usuarios_mediasComunales}
\end{figure}

 In Fig. \ref{Fig5_Usuarios_mediasComunales}, the median of the cosine similarity distributions between users and communities average media vectors are reported for the 2019 and 2020 data sets. Here, the comparison with their own communities are sketched in the diagonal, meanwhile the corresponding comparison with other communities can be found off the diagonal. These results show the consistency of these communities in terms of media consumer profiles. We observed that the main five communities can also be separated in two well differentiated groups: One dominated by {\it Clarin - La Nacion - Infobae}  and  the other one by  \textit{Pagina 12} and {\it El Destape}.

\par
\begin{figure}[h!]
 \centering
    \includegraphics[scale = .35]{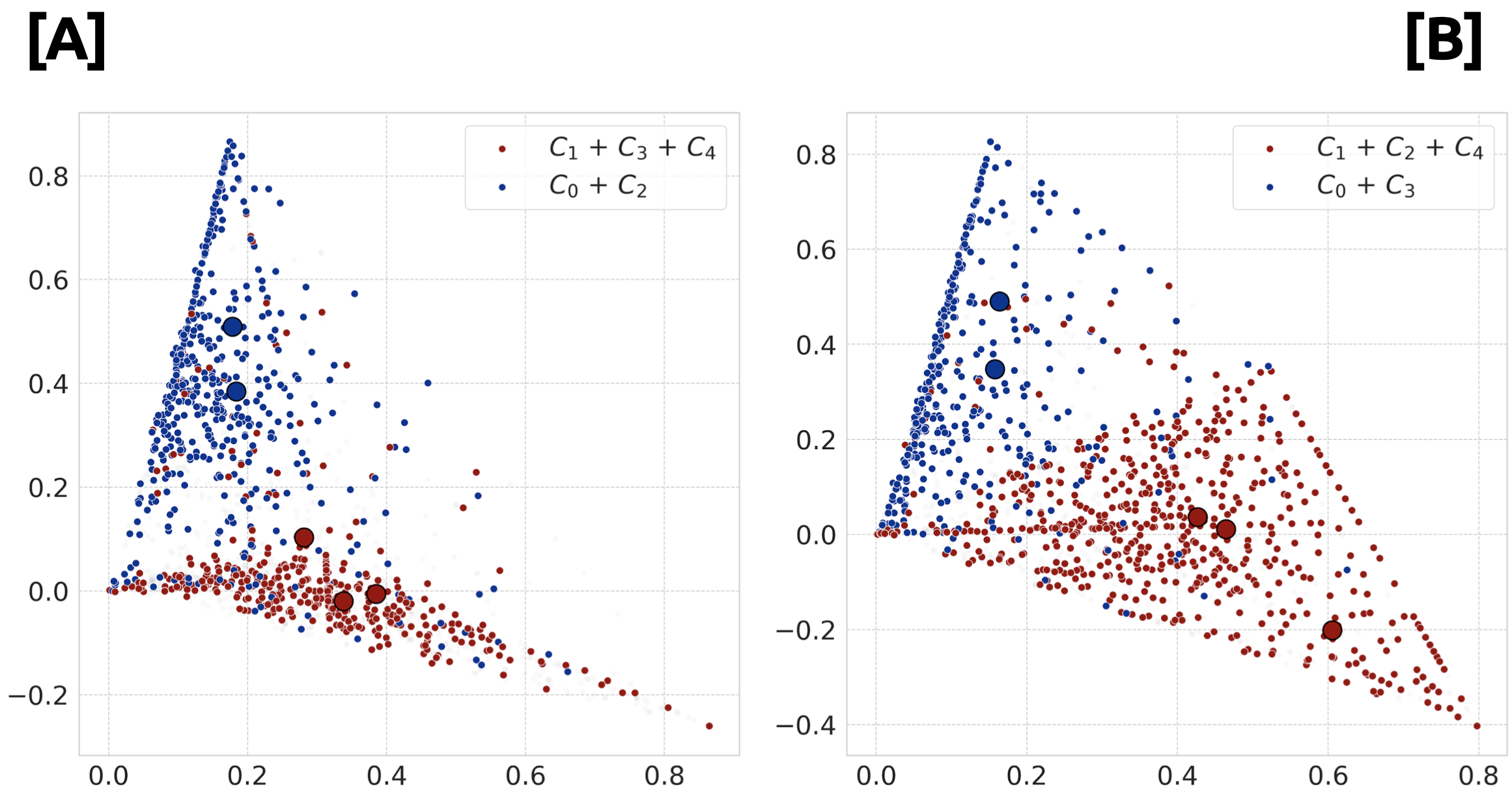}
    \caption{\textit{2020 and 2019 users corrected media vector mapping, after SVD transformation. Users belonging to communities identified previously as a block are coloured with the same color.}}
    \label{Fig6_Usuarios_pca}
\end{figure}

 In Fig. \ref{Fig6_Usuarios_pca}, average vectors for users and communities are embedded onto the first two dimensions of the Singular Value Decomposition (SVD) representation and colored according to communities. Here, we observe that the differences between the center-left and the center-right groups are not only at the level of average, but also between populations.

\par These results show that Argentinian media consumption in Twitter is polarized between two groups of users that reflect the political confrontation in National Elections between a center-left and a center-right political coalition. 

\subsubsection*{Diversity in the news consumption and their role in the communities}

\par Previous results show that the users tend to cluster in communities according to the media outlets they read or share. We analyzed here if the lack of diversity in media consumption (as defined in Eq. \ref{pol}) plays an important role in the formation of these communities. 

\par
\begin{figure}[h!]
     \centering
    \includegraphics[scale = .5]{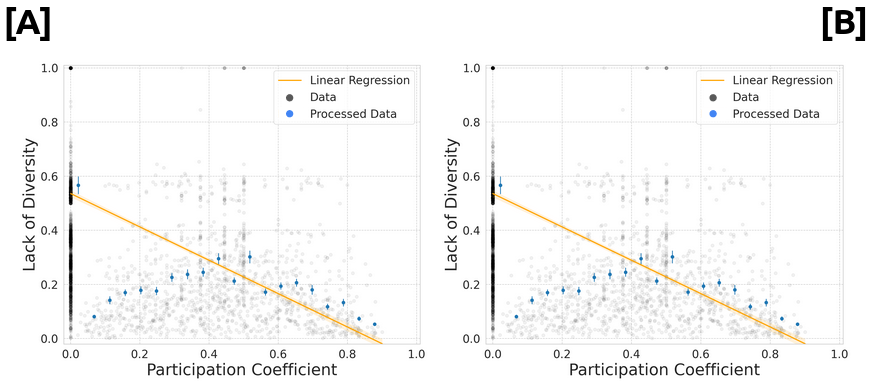}
    \caption{\textit{Lack of diversity vs participation coefficient. [A] and [B] display 2019 and 2020 data, respectively.}}
    \label{Fig7_cp_vs_pol}
\end{figure}

 In Fig. \ref{Fig7_cp_vs_pol}, we compare the lack of diversity in media consumption with the role of the users in the  networks. In particular, we chose the participation coefficient defined because its low values indicate a strong membership to a given module. 

\par Fig. \ref{Fig7_cp_vs_pol} shows that there is a negative correlation between the participation coefficient and the lack of diversity of the users in both data sets. The values of linear correlation for 2019 and 2020 are $-0.47$ and $-0.45$, respectively ($p_{val} < 0.01$ compared with a null model where node's $LD$ values were randomly shuffled). However, it is clear that the negative correlation is carried by those users that have participation coefficient (pc) $pc = 0$, while users with low but non-zero $pc$ present a different relationship with their $LD$. These results show that users that play a central role in each community have a noticeable lack of diversity in the news they consume. On the other hand, users that are in the border between communities tend to share news from different outlets.

\section*{Discussion}\label{Dis}

\par The role of social media in the formation of groups of like-minded users that share a given narrative (known as echo-chambers) has been extensively studied  given their profound impact in the formation of public opinion. 

\par The emergence of such echo chambers have been observed in networks defined by preferential message propagation (\emph{retweet networks}, in the case of Twitter) \cite{Menczer2011,Becatti2019} , as well as in networks of followers \cite{Cinelli2021} and mainly in discussions around a single topic.  They reflect the clustering of individuals based on different measures of similarity among users. For instance, in \cite{Cinelli2021} groups are identified by  the leaning of the media they tweet. In \cite{Vicario_2016, Zollo_2017}, according the ratio in pro and against likes on the discussed topic in Facebook, in \cite{Cota2019} they were defined by common hashtags and in \cite{Baronchelli_2021} by following similar groups of influencers.

\par It is also relevant to the analysis of ideological polarization processes where groups are formed by ideological affinity \cite{Bail2018}, which consists in sharing opinions on a group of issues \cite{Baumann_2021}.

\par Given the importance of the media in the processes of opinion formation \cite{McCombs1972,Guo2015} and how the affinity ties between users limit and constraint the circulation of information \cite{DelVicario554}, a relevant question is what is the role of the media in the emergence of these groups and which is the role of the  underlying political polarization in this process.

\par In this work we study if news sharing of Argentinian newspapers in social media produces  the emergence of homogeneous groups in terms of media consumption habits. Our main contribution  is the detection of echo chambers in bipartite user-news networks and their identification  in terms of consumption patterns of media outlets associated with the underlying ideological leaning patterns in Argentinean political life.

\par We selected two sets of groups of Twitter users and focused in tweets with links to media outlets they share.  With this data, we build a bipartite network of users and news and analyze their projections in both layers, focusing on the analysis of  emergent structures at a meso-scale level (communities) and the role of nodes in these structures. We compare the results between an electoral and non-electoral year and between a set of politically active users and a control group. 

\par The main result is the emergence of two main groups  of news articles identified by a unique distribution of media outlets in all analyzed datasets. On one side, news from 
{\it Página 12} and {\it El Destape} appears highly connected in one group (the center-lef one) and  news from {\it Clarín}, {\it La Nación} and {\it Infobae} in the other (center-right). This division is consistent with the ideological leaning of some of these outlets as was previously reported in \cite{mitchlstein2017, Becerra2012, Yeager2014}. The same distribution was found in the set of control users, indicating that these results do not depend if users show explicit political preferences. Figure S6 in Supplementary Information shows the similarities between these two groups across datasets.

\par We also verified if the ideological leaning of the media outlets parallels the one observed in the two main political Argentinian coalitions which disputed the national elections in 2019  by means of a sentiment analysis in the news articles belonging to the two main groups. The obtained values of the Sentiment Bias Statistics (SB) show consistently a significant difference between these two groups in both years. An interesting results was a change from a sentiment bias towards the candidates of the center-left coalition (or against the candidates of the center-right one) in 2019 to zero in 2020 in the groups of news articles dominated by {\it Página 12} and {\it El Destape}. A similar displacement in the values of the sentiment values from zero to a bias towards the candidates of the center-right coalition (or against the candidates of the center-left one) in the groups of news articles dominated by {\it Clarín}, {\it La Nación} and {\it Infobae}. A possible interpretation of these results is that in 2019, the candidates of the center-left coalition (Alberto Fernandez and Cristina Kirchner) won the national elections (and therefore a bias towards the winning candidates), meanwhile 2020 was a year dominated by the SARS-COV-2 pandemic and measures as prolonged quarantines was in detrimental of the incumbent president.

\par As expected, the analysis of the user projection network produced similar results. The news-sharing behavior give rise to the emergence of two defined groups of user which could be defined in terms of distribution of news from the media they share: One of them dominated by {\it Clarin - La Nacion - Infobae}  and  the other one by  \textit{Pagina 12} and {\it El Destape}.

\par Our results contribute to the characterization of echo chambers in terms of vector-media consumption and the use of sentiment bias to infer the leaning of media outlets. They also shed light on the process in which the political polarization in Argentina constrains the exposure to media content in social media. We should point out, however, that the sets of selected users, as well as the time span of their selected activity, clearly constraint the conclusions of our studies. Another limitation is the use of the Louvain algorithm to detect communities. Even though we check our results with other algorithms obtaining similar results, further analysis may include other alternatives, such as stochastic block model for instance. Future works should also expand these analysis to larger datasets (comprising longer periods of time) and to other countries. In particular, ideological leaning should be checked with different measures to test its robustness.

%%%%%%%%%%%%%%%%%%%%%%%%%%%%%%%%%%%%%%%%%%%%%%
%%                                          %%
%% Backmatter begins here                   %%
%%                                          %%
%%%%%%%%%%%%%%%%%%%%%%%%%%%%%%%%%%%%%%%%%%%%%%

\section*{Declarations}

\subsection*{Availability of data and materials}
    The datasets supporting the conclusions of this article are available in the GitHub repository,  in \href{https://github.com/LMDC-DF/twitter_mediaConsumption}{https://github.com/LMDC-DF/twitter\_mediaConsumption}
\subsection*{Competing interests}
  The authors declare that they have no competing interests.
\subsection*{Fundings}
 This work was supported by the Science and Technology Secretary, University of Buenos Aires (UBACyT), Argentina under Grant Number 20020130100582BA, and by the National Agency for Science and Technology Promotion (ANPCyT), Argentina, under Grant Number PICT-201-0215.
\subsection*{Author's contributions}
    TC and SMdP collected and analyzed the data. TC, SMdP, ET and PB interpret the results and wrote the manuscript. ET and PB conceived the study.
\subsection*{Acknowledgements}
  We thanks University of Buenos Aires and CONICET for supporting this research. We also thanks all members of SoPhy and CoCuCo Labs for fruitfully discussions.
  
%%%%%%%%%%%%%%%%%%%%%%%%%%%%%%%%%%%%%%%%%%%%%%%%%%%%%%%%%%%%%
%%                  The Bibliography                       %%
%%                                                         %%
%%  Bmc_mathpys.bst  will be used to                       %%
%%  create a .BBL file for submission.                     %%
%%  After submission of the .TEX file,                     %%
%%  you will be prompted to submit your .BBL file.         %%
%%                                                         %%
%%                                                         %%
%%  Note that the displayed Bibliography will not          %%
%%  necessarily be rendered by Latex exactly as specified  %%
%%  in the online Instructions for Authors.                %%
%%                                                         %%
%%%%%%%%%%%%%%%%%%%%%%%%%%%%%%%%%%%%%%%%%%%%%%%%%%%%%%%%%%%%%

% if your bibliography is in bibtex format, use those commands\bibliographystyle{bmc-mathphys} % Style BST file (bmc-mathphys, vancouver, spbasic).
%\bibliography{bmc_article}      % Bibliography file (usually '*.bib' )
% for author-year bibliography (bmc-mathphys or spbasic)
% a) write to bib file (bmc-mathphys only)
% @settings{label, options="nameyear"}
% b) uncomment next line
%\nocite{label}

% or include bibliography directly:
\bibliographystyle{unsrt}

\end{document}

\maketitle

\section{Keywords for the Twitter search of users}

In order to select both set of users (political engaged and those who mention media outlets), we use the list of keywords listed in Table \ref{list_keywords}. Those users who tweeted using some of this words where selected.

\begin{table}[h!]
\begin{center}
\begin{tabular}{ | p{0.17\linewidth}| p{0.78\linewidth} |} 
\hline
\textbf{User dataset} & \textbf{Keywords tweeted} \\ \hline
Politically active
users & Hashtag list: Elisacarrio, OfeFernandez$\_$, PatoBullrich, macri, macrismo, mauriciomacri, pichetto, MiguelPichetto, JuntosPorElCambio, alferdez, CFKArgentina, CFK, kirchner, kirchnerismo, FrenteTodos, FrenteDeTodos, Lavagna, RLavagna, Urtubey, UrtubeyJM, ConsensoFederal, 2030ConsensoFederal, DelCaño, NicolasdelCano, DelPla, RominaDelPla, FitUnidad, FdeIzquierda, Fte$\_$Izquierda, Castañeira, ManuelaC22, Mulhall, NuevoMas, Espert, jlespert, FrenteDespertar, Centurion, juanjomalvinas, Hotton, CynthiaHotton, Biondini, Venturino, FrentePatriota	, RomeroFeris, PartidoAutonomistaNacional, Vidal, mariuvidal, Kicillof, Kicillofok, Bucca, BuccaBali, chipicastillo, Larreta, horaciorlarreta, Lammens, MatiasLammens, Tombolini, matiastombolini, Solano, Solanopo, Lousteau, GugaLusto , Recalde, marianorecalde , RAMIROMARRA, Maxiferraro , fernandosolanas , MarcoLavagna , myriambregman , cristianritondo, Massa, SergioMassa , GracielaCamano, nestorpitrola \\

\hline 
Users control group &  La Nación, Clarín, Infobae, Página 12, El Destape, La Izquierda Diario, Ámbito Financiero, Radio Mitre, Minuto Neuquén, Rosario 3, El Cronista, El Chubut \\ \hline
\end{tabular}
\caption{\textit{Keyword lists used to select users.}}
\label{list_keywords}
\end{center}
\end{table}

\section{Bipartite Degree for News and Users}
 %As mentioned in Methods, an hyperbolic projection of the bipartite network was used. To justify this choice, 
\par
\begin{figure}[h!]
    \centering
    \captionsetup{justification=centering,margin=0.01cm}    
    \includegraphics[scale = .5]{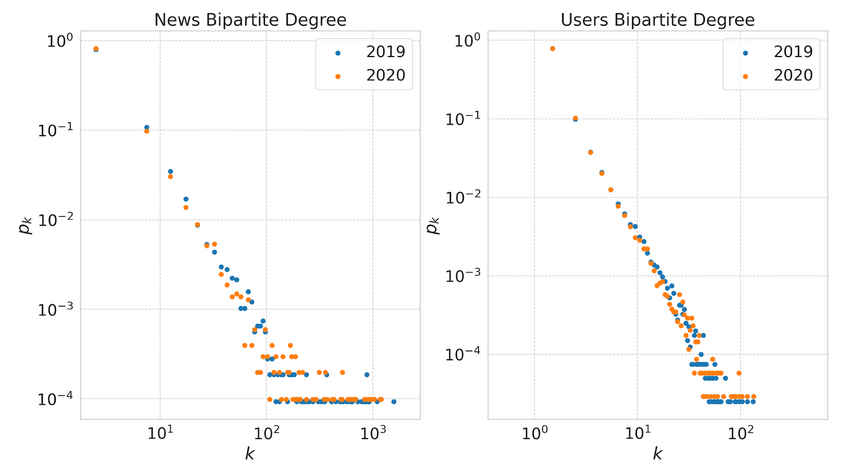}
    \caption{\textit{Degree of news and users in the bipartite networks. On the one hand, the user degree refers to the number of news shared by the user; on the other hand, news degree stands for the number of users that shared a new.}}
    \label{fig_s1}
\end{figure}

 The  degree distributions of the bipartite networks for users and news are shown in Fig. \ref{fig_s1}. As it can be seen, the long tail distributions implies that there are users that share compulsively news, as well as there are news that are consumed multiple times. Therefore, an hyperbolic projection turns naturally, to mitigate the effects of this highly connected nodes.

\section{Summary of Networks Topology}
\par In Table \ref{topology_summary}, we provide a summary of different topological properties of all analyzed networks. Although the amount of users selected in both datasets was the same, after discarding tweets that do not contain media outlet URLs, differences between them can be seen. Networks from politically active dataset are larger and less dense than those from control group. On one hand, control group news networks present less edges (E) and mean degree ($<k>$) than the politically active networks. On the other hand, control group users networks have more edges and $<k>$ than politically active ones.

\begin{table}[h!]
\begin{tabular}{|l|cccc|cccc|}
\hline
\multirow{2}{*}{}                                                     & \multicolumn{4}{c|}{\textbf{Users Networks}}                                                                                 & \multicolumn{4}{c|}{\textbf{News Networks}}                                                                                  \\ \cline{2-9} 
                                                                      & \multicolumn{2}{c|}{\textbf{Politically active}}                        & \multicolumn{2}{c|}{\textbf{Control Group}}        & \multicolumn{2}{c|}{\textbf{Politically active}}                        & \multicolumn{2}{c|}{\textbf{Control Group}}        \\ \hline
\textbf{Year}                                                         & \multicolumn{1}{c|}{\textbf{2020}} & \multicolumn{1}{c|}{\textbf{2019}} & \multicolumn{1}{c|}{\textbf{2020}} & \textbf{2019} & \multicolumn{1}{c|}{\textbf{2020}} & \multicolumn{1}{c|}{\textbf{2019}} & \multicolumn{1}{c|}{\textbf{2020}} & \textbf{2019} \\ \hline
\textbf{N}                                                            & \multicolumn{1}{c|}{4323}          & \multicolumn{1}{c|}{3625}          & \multicolumn{1}{c|}{3021}          & 1844          & \multicolumn{1}{c|}{10975}         & \multicolumn{1}{c|}{12221}         & \multicolumn{1}{c|}{6597}          & 4734          \\ \hline
\textbf{E}                                                            & \multicolumn{1}{c|}{15621}         & \multicolumn{1}{c|}{10154}         & \multicolumn{1}{c|}{18603}         & 5718          & \multicolumn{1}{c|}{83611}         & \multicolumn{1}{c|}{111422}        & \multicolumn{1}{c|}{22140}         & 12283         \\ \hline
\textbf{\textless{}k\textgreater{}}                                   & \multicolumn{1}{c|}{7.23}          & \multicolumn{1}{c|}{5.60}          & \multicolumn{1}{c|}{12.32}         & 6.20          & \multicolumn{1}{c|}{15.24}         & \multicolumn{1}{c|}{18.23}         & \multicolumn{1}{c|}{6.71}          & 5.19          \\ \hline
\textbf{\textless{}s\textgreater{}}                                   & \multicolumn{1}{c|}{3.38}          & \multicolumn{1}{c|}{3.99}          & \multicolumn{1}{c|}{3.10}          & 2.85          & \multicolumn{1}{c|}{1.36}          & \multicolumn{1}{c|}{1.34}          & \multicolumn{1}{c|}{1.35}          & 1.17          \\ \hline
\textbf{\begin{tabular}[c]{@{}l@{}}Average\\ Clustering\end{tabular}} & \multicolumn{1}{c|}{0.30}          & \multicolumn{1}{c|}{0.24}          & \multicolumn{1}{c|}{0.31}          & 0.25          & \multicolumn{1}{c|}{0.51}          & \multicolumn{1}{c|}{0.52}          & \multicolumn{1}{c|}{0.40}          & 0.35          \\ \hline
\textbf{Density}                                                      & \multicolumn{1}{c|}{0.0017}        & \multicolumn{1}{c|}{0.0015}        & \multicolumn{1}{c|}{0.0041}        & 0.0034        & \multicolumn{1}{c|}{0.0014}        & \multicolumn{1}{c|}{0.0015}        & \multicolumn{1}{c|}{0.0010}        & 0.0011        \\ \hline
\end{tabular}
\caption{\textit{Summary of the topological properties of the networks from both data sets. PASO and Control Group refer to the political engaged users data set and the users that mentioned media outlets data set, respectively.}}
\label{topology_summary}
\end{table}

\section{Normalized Mutual Information between Partitions}
\par In order to compare communities obtained from Louvain community detection algorithm, normalized mutual information was performed between the 20 main communities from partitions detected by three different methods: Louvain \cite{louvain}, Infomap \cite{infomap} and Label Propagation \cite{labelprop}. The results of this analysis are shown in Table \ref{nmis}. As it can be seen, the detected communities are similar in terms of normalized mutual information in all cases.

\begin{table}[h!]
\begin{tabular}{|ll|ll|llc|}
\hline
\multicolumn{2}{|l|}{\multirow{2}{*}{\textbf{News Networks}}}                     & \multicolumn{2}{c|}{\textbf{2019}}                                            & \multicolumn{3}{c|}{\textbf{2020}}                                                                             \\ \cline{3-7} 
\multicolumn{2}{|l|}{}                                                            & \multicolumn{1}{c|}{\textbf{Louvain}} & \multicolumn{1}{c|}{\textbf{Infomap}} & \multicolumn{1}{c|}{\textbf{Infomap}} & \multicolumn{1}{c|}{\textbf{Louvain}} &                                \\ \hline
\multicolumn{1}{|l|}{\multirow{2}{*}{\textbf{2019}}} & \textbf{Label propagation} & \multicolumn{1}{l|}{0.83}             & 0.78                                  & \multicolumn{1}{l|}{0.91}             & \multicolumn{1}{l|}{0.93}             & \multirow{2}{*}{\textbf{2020}} \\ \cline{2-6}
\multicolumn{1}{|l|}{}                               & \textbf{Infomap}           & \multicolumn{1}{l|}{0.92}             & -                                     & \multicolumn{1}{l|}{-}                & \multicolumn{1}{l|}{0.87}             &                                \\ \hline
\multicolumn{2}{|l|}{\multirow{2}{*}{\textbf{Users Networks}}}                    & \multicolumn{2}{c|}{\textbf{2019}}                                            & \multicolumn{2}{c|}{\textbf{2020}}                                            &                                \\ \cline{3-7} 
\multicolumn{2}{|l|}{}                                                            & \multicolumn{1}{c|}{\textbf{Louvain}} & \multicolumn{1}{c|}{\textbf{Infomap}} & \multicolumn{1}{c|}{\textbf{Infomap}} & \multicolumn{1}{c|}{\textbf{Louvain}} &                                \\ \hline
\multicolumn{1}{|l|}{\multirow{2}{*}{\textbf{2019}}} & \textbf{Label propagation} & \multicolumn{1}{l|}{0.70}             & 0.76                                  & \multicolumn{1}{l|}{0.82}             & \multicolumn{1}{l|}{0.89}             & \multirow{2}{*}{\textbf{2020}} \\ \cline{2-6}
\multicolumn{1}{|l|}{}                               & \textbf{Infomap}           & \multicolumn{1}{l|}{0.89}             & -                                     & \multicolumn{1}{l|}{-}                & \multicolumn{1}{l|}{0.82}             &                                \\ \hline
\end{tabular}
\caption{\textit{Normalized mutual information score between the partitions of the 20 main communities obtained from applying Louvain, Infomap and Label Propagation algorithms on both, users and news, networks.}}
\label{nmis}
\end{table}

\section{Topic Description}
\par In the main work, a topic decomposition of the news content in the set of politically active users was performed. In this section, word clouds for each topic of the main two communities of news for both years are shown. It should be noticed that in both 2019 communities, a topic related with the \textbf{National Election} emerge. Among the most frequent words (in terms of tf-idf) appears the names of the candidates of the two main political coalitions on the electoral dispute. This co-appearance aims us to compute the sentiment bias, as mentioned on the main work.

\begin{figure}[h!]
    \centering
    \captionsetup{justification=centering,margin=0.01cm}    
    \includegraphics[scale = .45]{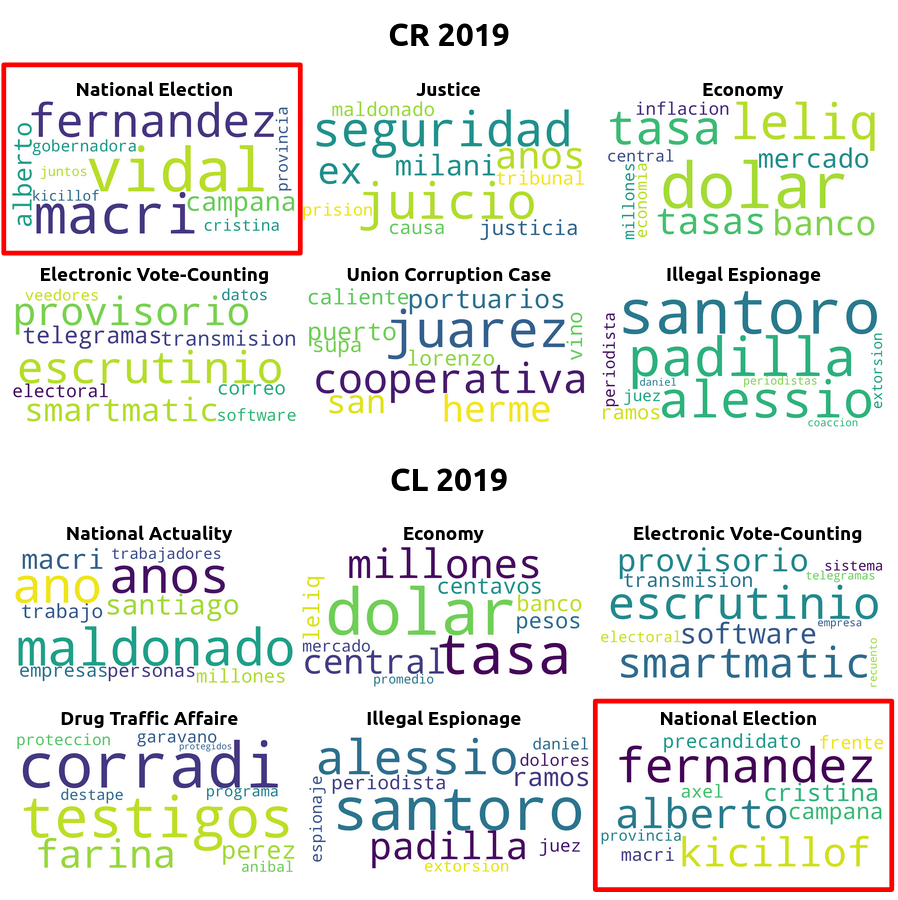}
    \caption{\textit{Topic description on the term space, mapped on word clouds, where the size of each word is proportional to the weight of the word on each topic. Center right and center left main communities of 2019 news network are shown. Word clouds boxed in red highlight the National Election topics of both communities, were candidate names appear.}}
    \label{fig_s2_1}
\end{figure}
\begin{figure}[h!]
    \centering
    \captionsetup{justification=centering,margin=0.01cm}    
    \includegraphics[scale = .45]{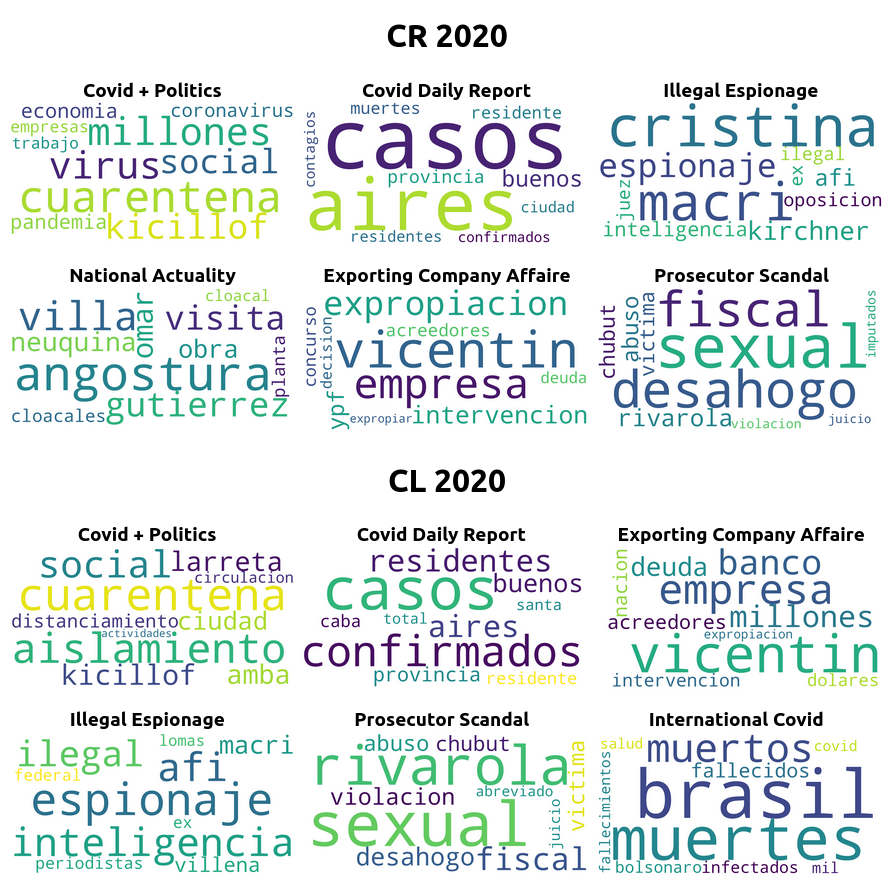}
    \caption{\textit{Topic description on the term space, mapped on word clouds, where te size of each word is proportional to the weight of the word on each topic. Center right and center left main communities of 2020 news network are shown.}}
    \label{fig_s2_2}
\end{figure}

\section{Analysis of the Control Data Set}
\par As mentioned in the main work, a control data set was analyzed in order to asses the robustness of our original results. Analysis for both, users and news networks, can be found on the next subsections.

\subsection{The news projection}

\par
\begin{figure}[h!]
    \captionsetup{justification=centering,margin=0.01cm}    
    \includegraphics[scale = .35]{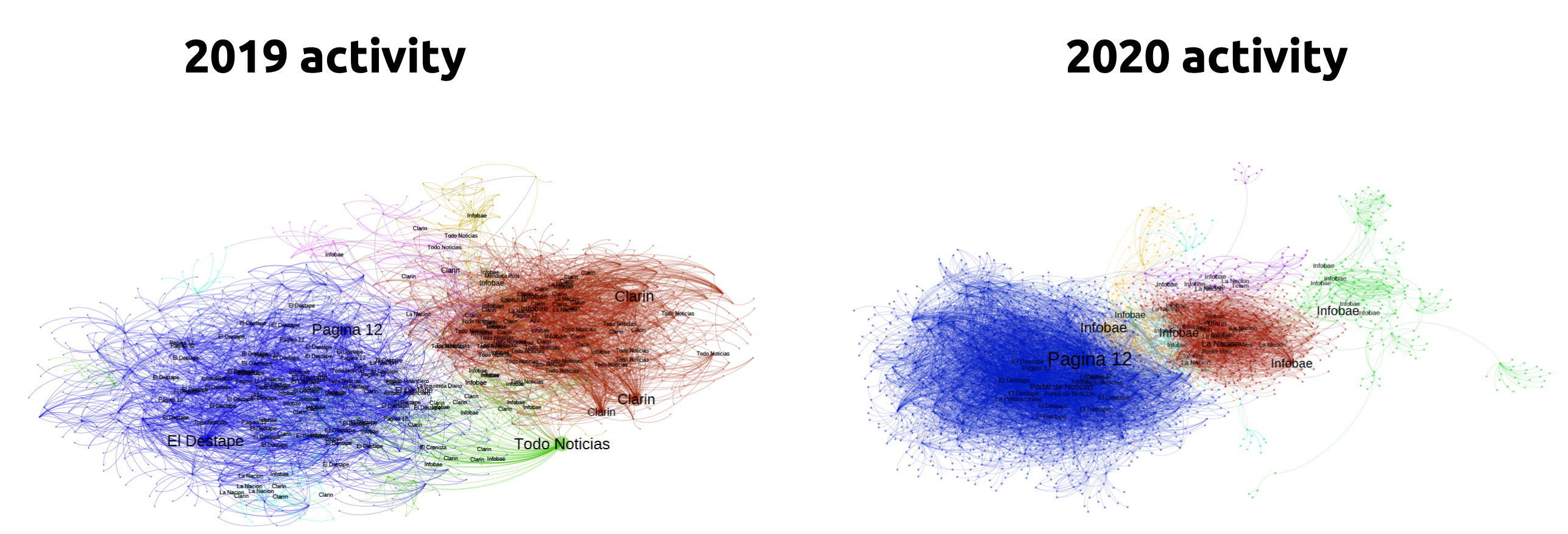}
    \caption{\textit{2019 and 2020 news networks visualization. Nodes were colored by community membership and labeled by media outlet proportionally to their within-module degree}.}
    \label{fig_s3}
\end{figure}

 Firstly, we present the results of the analysis of the news projection of the control group data set. In Fig. \ref{fig_s3}, a visualization of both, 2019 and 2020, networks are shown. As before, two main communities emerge, were each one is dominated by a given group of media outlets: on the one hand, \textit{Pagina 12} and \textit{El Destape}; and on the other hand, \textit{Clarin}, \textit{Infobae} and \textit{La Nación}. 
\par
\begin{figure}[h!]
    \centering
    \captionsetup{justification=centering,margin=0.01cm}    
    \includegraphics[scale = .5]{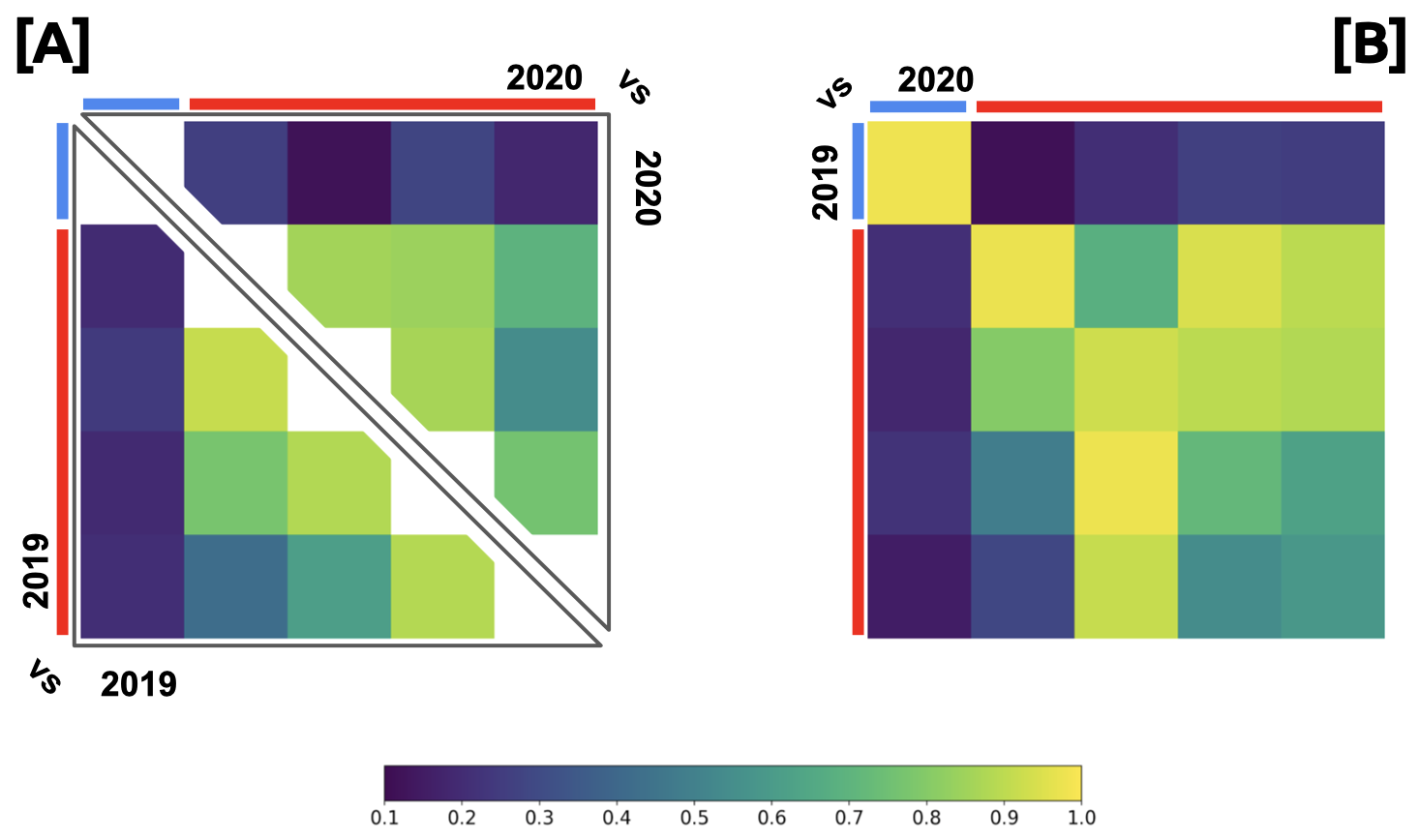}
    \caption{\textit{Consistency of news network. Panel [A] accounts for similarities among communities of same years and Panel [B] compare 2019 against 2020}.}
    \label{fig_s4}
\end{figure}

 Motivated by the previous network visualizations, we compute the cosine similarity between the media outlet distribution of the main communities. This similarity was computed between communities of the same year and also between communities of different years. As it is shown in Fig. \ref{fig_s4}, it is possible to identify two groups of communities: the Center-Right ones and the Center-Left ones.
\par
\begin{figure}[h!]
    \centering
    \captionsetup{justification=centering,margin=0.01cm}    
    \includegraphics[scale = .5]{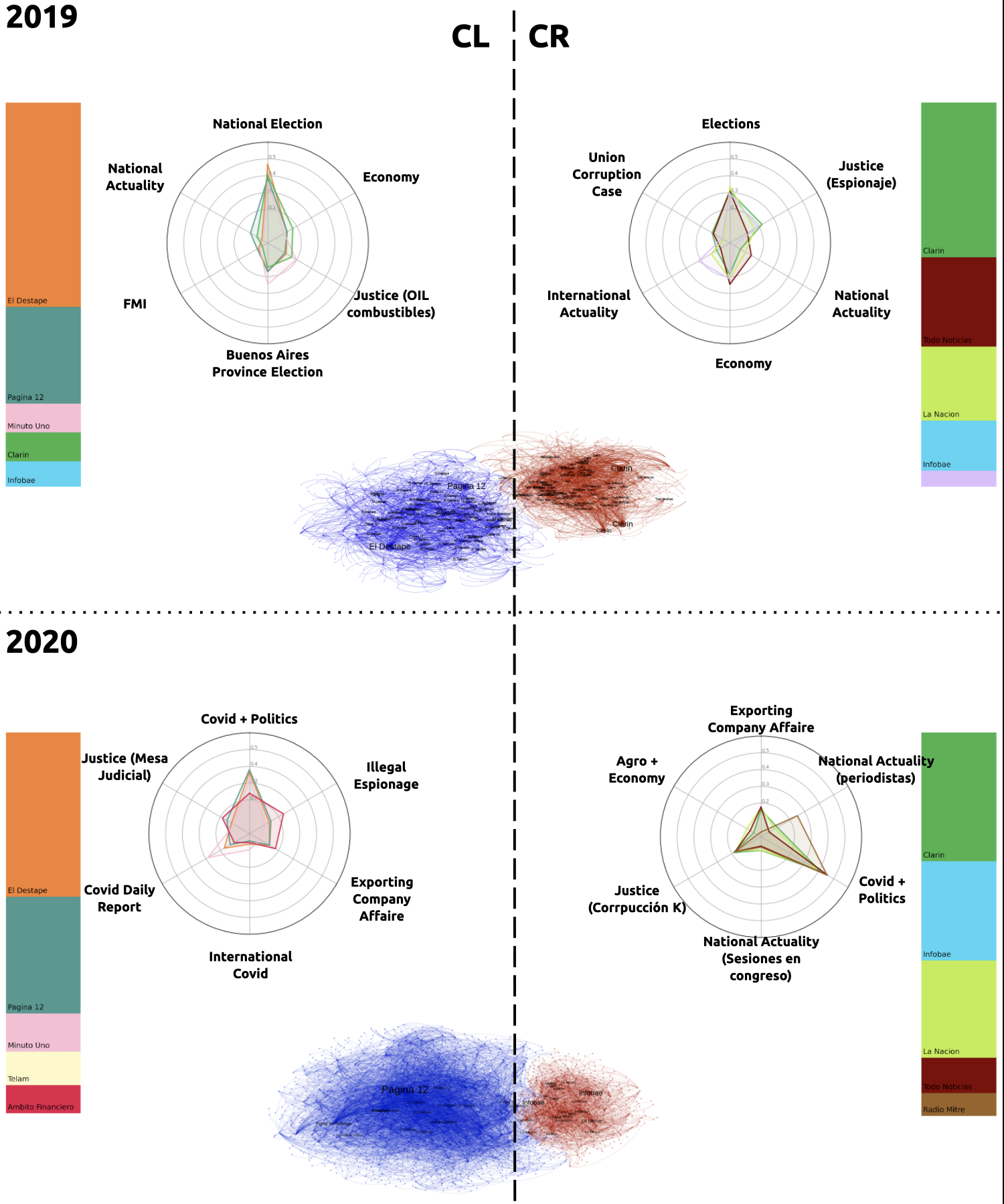}
    \caption{\textit{Media outlets distributions and topic decomposition for the 2019 ans 2020 two main communities. The stacked bars represents the media outlet distribution, while the radar plot  display the media agenda. The different colored lines in radar plot indicate the agendas of each outlet with the same color as in the stacked bar.}}
    \label{fig_s5}
\end{figure}

 Then, for the main two communities of each year, a media outlet distribution and a topic decomposition of the news content analysis were performed. Results are presented in Fig. \ref{fig_s5}. It is notorious that both communities are dominated by different set of media outlets, as highlighted before.

\begin{figure}[h!]
    \centering
    \captionsetup{justification=centering,margin=0.01cm}    
    \includegraphics[scale = .4]{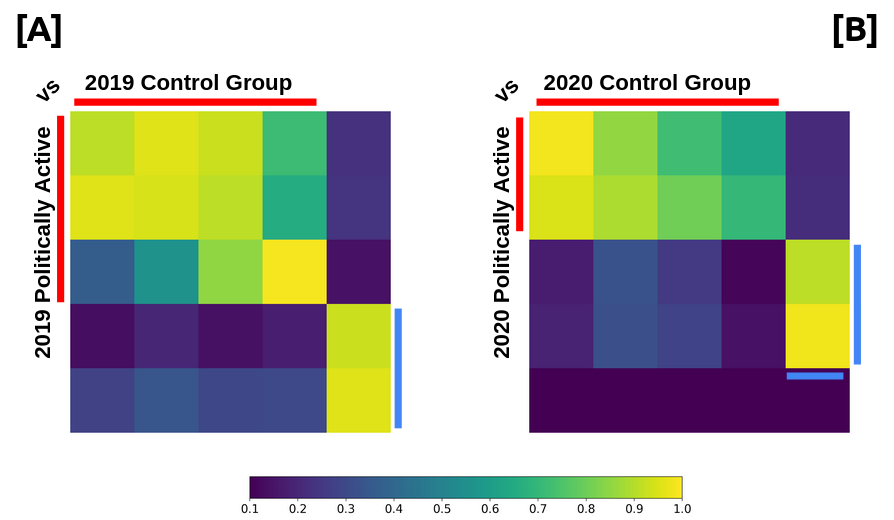}
    \caption{\textit{Cosine similarities between the main five communities of the politically active and control group news networks. 2019 networks are compared on panel [A], while 2020 similarities are shown on panel [B].}}
    \label{fig_s6}
\end{figure}

Finally, we decide to compare the media outlet distributions between the main five communities of the control group news networks and the main five communities of the politically active news networks. The computed cosine similarities are shown in Fig. \ref{fig_s6}

We cab appreciate here the high degree of similarity in center-left and center-right groups between both datasets.

\subsection{The users projection}
\par
\begin{figure}[h!]
    \centering
    \captionsetup{justification=centering,margin=0.01cm}    
    \includegraphics[scale = .5]{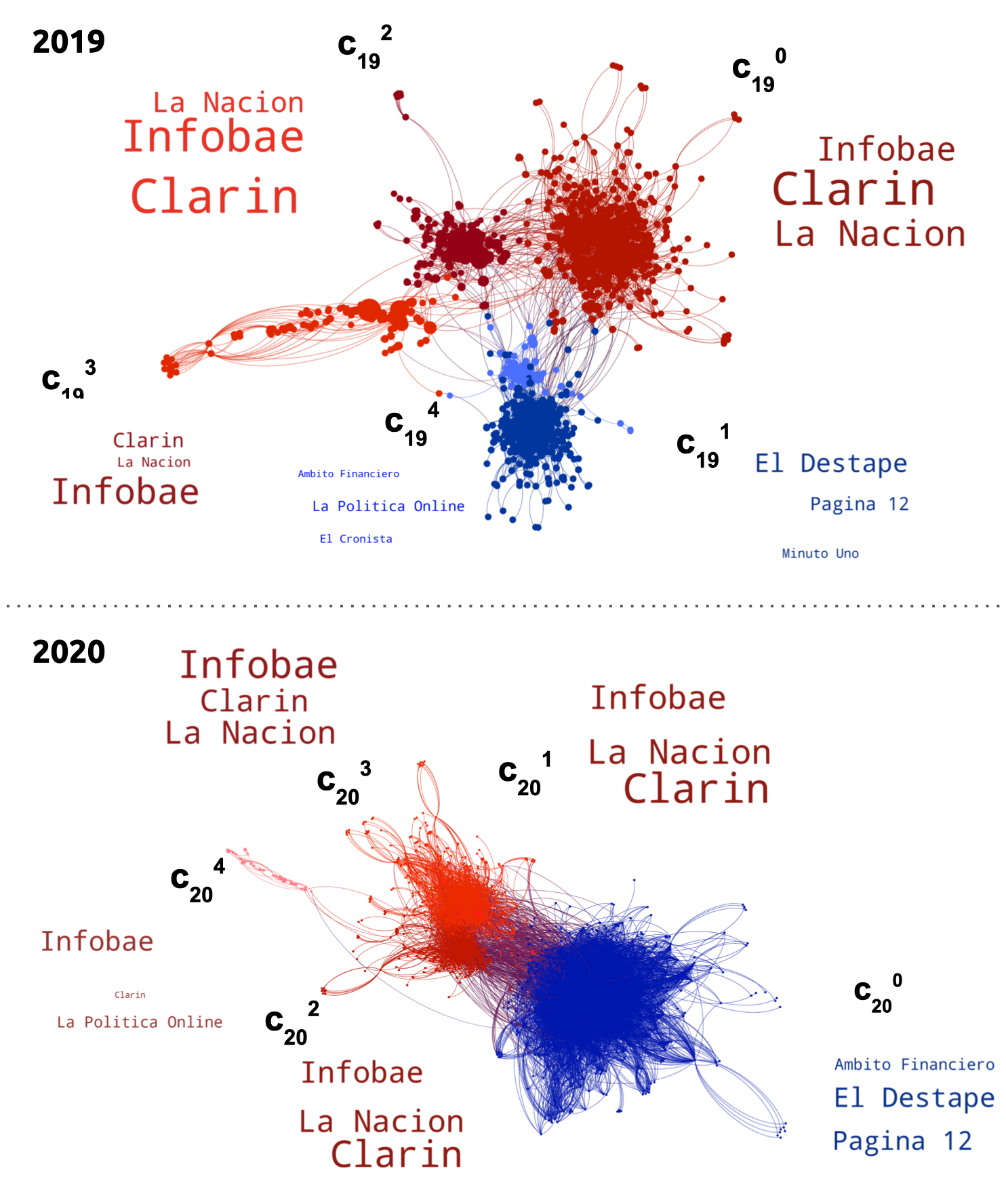}
    \caption{\textit{2019 and 2020 user networks visualization. Nodes were colored by communities membership. Word clouds display the averaged corrected media distribution of each community.}}
    \label{fig_s7}
\end{figure}

 Now, we present the results of the analysis of the users projection of the control group data set. In Fig. \ref{fig_s7}, a visualization of 2019 and 2020 networks are shown, along with the corresponding word clouds that represent the average media-consumed vector of the main communities. Communities where colored by users news consumption. Those communities in which users tweeted mainly with links to center-left media outlets were colored in red and those with links to center-right outlets in blue.

\par
\begin{figure}[h!]
    \centering
    \captionsetup{justification=centering,margin=0.01cm}    
    \includegraphics[scale = .5]{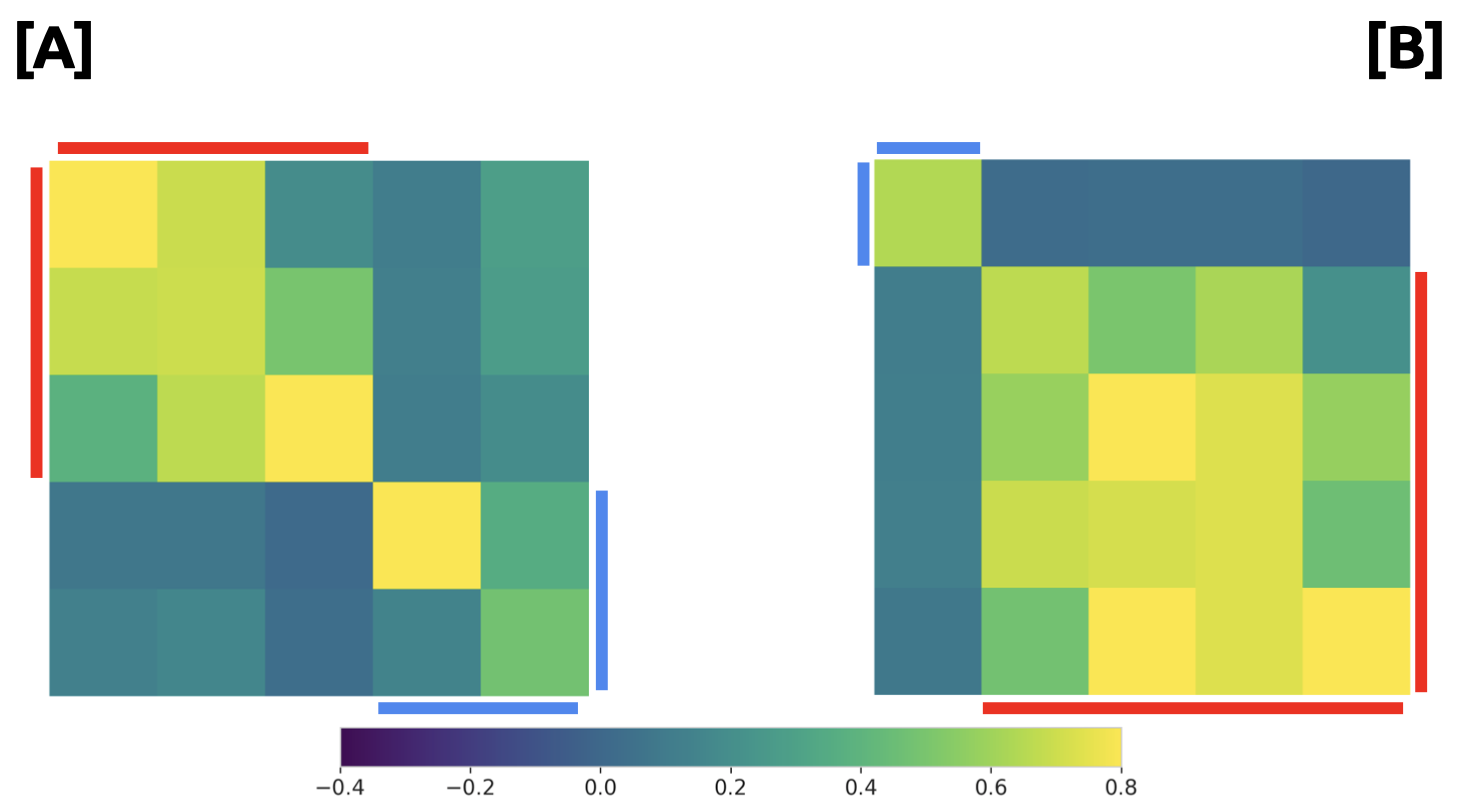}
    \caption{\textit{Similarities between users and average communities in media-consumed vectors. [A] and [B] accounts for 2019 and 2020 data sets, respectively. The i,j-th element of each figure corresponds to compute de median of the cosine similarities distribution between the average media-consumed vector of the i-th community and all the users media-consumed vectors belonging to the j-th community.}}
    \label{fig_s8}
\end{figure}

 In Fig. \ref{fig_s8}, we compute the similarities between the average media-consumed vector of the main communities and the media-consumed vector of the users, for both years. As seen in the main analysis, a group of Center-Right and a group of Center-Left communities emerge.
\par
\begin{figure}[h!]
    \centering
    \captionsetup{justification=centering,margin=0.01cm}    
    \includegraphics[scale = .5]{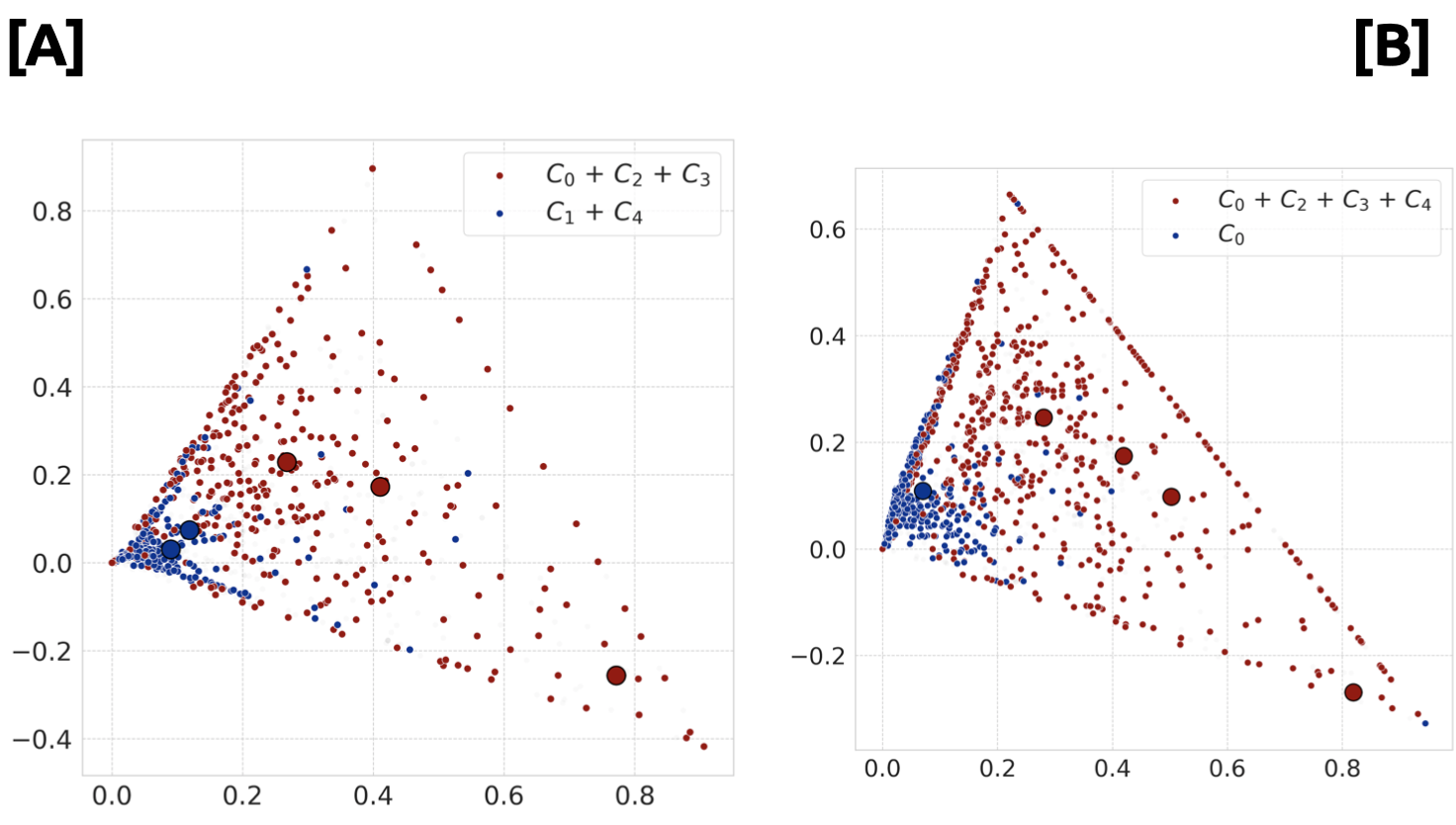}
    \caption{\textit{2020 and 2019 users corrected media vector mapping, after SVD transformation. Users belonging to communities identified previously as a block are coloured with the same color.}}
    \label{fig_s9}
\end{figure}

 And finally, the 2020 and 2019 users corrected media vector mapping is shown in Fig. \ref{fig_s9}. Those users previously detected in the same community are colored with the same color, red for center-right community and blue for center- community. It can be seen a similarly communities structure than  politically active users network has. Also there is a clear difference in media consumption depending on the user communities

\clearpage

\bibliographystyle{unsrt}